\documentclass[dvipsnames]{mr}

\usepackage{lineno}         
\usepackage{amsmath}        
\usepackage{amsfonts}       
\usepackage{nicefrac}       
\usepackage{xcolor}         
\usepackage{pgfplots}       
\usepackage{caption}        
\usepackage{arydshln}       
\usepackage{float}          
\pgfplotsset{compat=1.18}   

\title{\Large TASTE2: Text-Aligned Speech Modeling\\and Deployment toward Full-Duplex Voice Interaction}

\author[*1]{Yi-Chang Chen}
\author[*23]{Chun Wei Chen}
\author[*23]{Dien-Ruei Wu}
\author[3]{Jie Lin}
\author[4]{Yu-Kuan Fu}
\author[4]{Yang-Hsien Lin}
\author[4]{Eddie TC Huang}
\author[5]{Simon See}
\author[3]{Hung-yi Lee}
\author[1]{Da-Shan Shiu}

\contribution[*]{Equal contribution}

\affiliation[1]{MediaTek Research}
\affiliation[2]{Internship at MediaTek Research}
\affiliation[3]{National Taiwan University}
\affiliation[4]{NVAITC, NVIDIA, Taiwan}
\affiliation[5]{NVAITC, NVIDIA, Santa Clara, USA}

\metadata[Contact]{\texttt{\{yi-chang.chen, wilson-cw.chen, dienruei.wu, ds.shiu\}@mtkresearch.com, \{b11705048, hungyilee\}@ntu.edu.tw}}

\abstract{
Full-duplex voice interaction requires more than converting one complete utterance into another. A system must process speech as it arrives, decide when to take or yield the floor, and stop when the user interrupts, while preserving pretrained linguistic competence and acoustic paralinguistic cues. We ask whether TASTE (Text-Aligned Speech Tokenization and Embedding) provides a viable modeling path toward this goal.

We present TASTE2, which transforms the original utterance level TASTE method into an incremental dialogue stack. A shared text-token vocabulary removes word-level averaging across the Speech Tokenizer, Spoken LM, and Speech Detokenizer, while modality aligned dialogue training predicts one continuous audio latent for each text token without interleaving heterogeneous token streams. Furthermore, an incremental Speech Detokenizer enables streaming synthesis through CosyVoice2. After speech and dialogue training, TASTE2 (Merge) reaches 56.3\% on LLaMA-Questions against a 57.3\% Qwen2.5-7B Instruct text-only reference (98.2\% accuracy retention), and TASTE2 (Direct) reaches 53.0\% (92.4\% retention).

We also move beyond model level experiments by building TASTE2 VoiceBot, a working research system that processes user speech incrementally, streams synthesized audio, and stops generation when the user barges in. On Full-Duplex-Bench v1.0, TASTE2 and TASTE2 VoiceBot handle user interruptions well while maintaining high conversational coherence. Natural conversation conditions remain challenging, and the deployed mean time to first audio remains 2.701 s on two NVIDIA RTX A6000 after TensorRT acceleration.

Finally, to our knowledge, we provide the first systematic characterization of explicit paralinguistic control in a TASTE based model. Fast speaking rate serves as a cross-strategy proof of concept after dialogue SFT, while emotion control is strategy dependent and the remaining attributes stay weak. Together, these results establish TASTE based modeling as a practical route toward full-duplex systems while identifying natural conversation robustness, speech generation latency, and feature general paralinguistic control as open challenges. Explore TASTE2 online.\footnote{\textbf{TASTE2 resources:} \href{https://gitycc.github.io/TASTE2-Homepage/}{Home page} $\diamond$ \href{https://github.com/GitYCC/TASTE-SpokenLM-2}{Spoken LM implementation} $\diamond$ \href{https://github.com/GitYCC/TASTE-Voice-Bot}{VoiceBot deployment stack} $\diamond$ \href{https://huggingface.co/collections/YC-Chen/taste2}{Pretrained checkpoints} $\diamond$ \href{https://huggingface.co/datasets/wilzzzz/paralinguistic_dialogues}{Paralinguistic dialogue corpus}.}
}

\begin{document}

\maketitle

\newpage
\tableofcontents
\newpage

\section{Introduction}
\label{sec:introduction}

Natural voice interaction extends beyond mapping an utterance to a response. A conversational system must interpret linguistic content and paralinguistic cues, detect when to take the floor, and stop or revise its response when the user interrupts. Full-duplex dialogue systems such as Moshi~\citep{moshi}, SALMONN-omni~\citep{salmonn-omni}, and Freeze-Omni~\citep{freeze-omni} pursue these capabilities through continuous bidirectional processing, including turn taking, backchanneling, and barge-in handling, and are evaluated with benchmarks such as Full-Duplex-Bench~\citep{full-duplex-bench}. Two tensions make this goal difficult.

First, adding speech representations can weaken the linguistic competence inherited from a pretrained language model~\citep{think-before-talk,moshi}. Text pretraining establishes regularities among text tokens. Interleaving those tokens with acoustically oriented codes forces the model to represent linguistic content and acoustic realization in the same sequence, which can disrupt the patterns learned during pretraining. Existing systems mitigate this effect with Moshi's inner-monologue decoding, Freeze-Omni's multitask training, or large-scale joint training, but preserving text-model competence remains an open problem. The cost scales with how much of the sequence acoustic information occupies. Specifically, the parallel decoding streams of Mini-Omni~\citep{mini-omni} retain only 29.7\% of the text model's benchmark accuracy, and Moshi's multiple codebook streams retain 83.1\%, whereas TASTE2, whose sequence stays as long as the text sequence, retains 98.2\% (\S\ref{subsec:semantic_preservation}).

Second, system integration trades modularity against information preservation. End-to-end models jointly optimize speech understanding, response generation, and synthesis~\citep{glm4-voice,salmonn-omni}, allowing paralinguistic information to flow across the model but coupling the components operationally. Modular ASR--LLM--TTS pipelines allow each component to be improved independently, yet their text-only interfaces discard prosody, emotion, and speaking style before these cues can influence the response~\citep{paras2s,emo-reasoning}. A practical architecture should retain relevant acoustic information without giving up modular deployment.

TASTE (Text-Aligned Speech Tokenization and Embedding)~\citep{taste} offers a useful starting point. It aligns one speech representation with each text token, separating linguistic content from acoustic realization while avoiding the sequence-length mismatch common to joint text--speech modeling. Its original implementation, however, is not designed for incremental dialogue. The Whisper-based Speech Tokenizer and LLaMA tokenizer use different vocabularies, so TASTE averages representations over word-level boundaries. This operation can discard fine-grained information, requires language dependent segmentation, and complicates dialogue templates. Moreover, its synthesis stack operates on complete utterances rather than incrementally. TASTE also did not evaluate whether its aligned representations support turn-taking, interruption handling, or explicit paralinguistic control.

We present TASTE2, a model family and deployment architecture that establish text-aligned speech modeling as a viable path toward full-duplex voice interaction. The central representation contains two aligned information streams. Discrete \emph{text tokens} carry linguistic content while continuous \emph{audio latents} carry acoustic information. TASTE2 preserves this distinction throughout tokenization, language modeling, and synthesis. We make five contributions:

\begin{itemize}
    \item \textbf{An end-to-end adaptation of TASTE into an incremental dialogue stack.} A shared text-token vocabulary removes word-level averaging, modality-aligned dialogue training predicts one continuous audio latent per text token without interleaving heterogeneous token streams, and an incremental Speech Detokenizer enables streaming synthesis.
    \item \textbf{Semantic preservation without sequence inflation.}  By keeping the sequence strictly at text length, TASTE2(Merge) successfully retains the vast majority of its underlying text only reasoning accuracy. Furthermore, analyzing TASTE2(Direct) reveals that performance limits stem from instruction tuning integration rather than the continuous speech modeling itself.
    \item \textbf{To our knowledge, the first systematic characterization of explicit paralinguistic control in a TASTE-based model.} Fast speaking rate serves as a cross-strategy proof of concept after dialogue SFT, while emotion control is strategy dependent and the remaining attributes stay weak.
    \item \textbf{A working deployment.} TASTE2 VoiceBot processes user speech incrementally, decides when to take the floor, streams synthesized audio, and stops when the user barges in.
    \item \textbf{Full duplex conversational ability.} TASTE2 and TASTE2 VoiceBot  demonstrates immediate responsiveness to user interruptions on Full-Duplex-Bench v1.0 while maintaining high conversational coherence afterward.
\end{itemize}

Together, these contributions move TASTE from utterance level spoken language modeling to an implemented dialogue system and provide evidence that its aligned representation is a practical foundation for further full-duplex development. They also outline what remains to be solved, as active overlap behaviors are not yet deployed, natural conversation interaction is not uniformly strong, speech generation latency remains substantial, and explicit paralinguistic control is not consistent across attributes.

\section{Related Work}
\label{sec:related_work}

TASTE2 connects text-aligned speech modeling with incremental dialogue deployment. We review speech tokenization and joint text speech modeling (Section~\ref{ssec:rw_slm}), full-duplex architectures and turn taking (Section~\ref{ssec:rw_duplex}), and paralinguistic control and adaptation (Section~\ref{ssec:rw_para}).

\subsection{Speech Tokenization and Spoken Language Models}
\label{ssec:rw_slm}

\subsubsection{From Cascades to Joint Modeling}
Traditional ASR--LLM--TTS cascades allow each component to be optimized independently. Because only text crosses their interfaces, however, paralinguistic cues such as emotion, prosody, and speaking rate cannot influence the language model~\citep{voicebench,audiobench}.
GSLM~\citep{gslm, dgslm} showed that language models can instead generate coherent speech directly from discrete self-supervised units, and AudioLM~\citep{audiolm} extended this to long-form audio continuation by cascading semantic and acoustic token streams.
Subsequent work grafted speech onto pretrained text LLMs.
SpeechGPT~\citep{speechgpt} and AudioPaLM~\citep{audiopalm} did so through cross-modal instruction tuning, while Qwen-Audio~\citep{qwen-audio} and SALMONN~\citep{salmonn} used encoder-adapter interfaces.
These efforts established joint text-speech modeling as the dominant paradigm, though whether the resulting models preserve the semantic competence of the underlying text model remains the central open question.

\subsubsection{Speech Tokenization}
The representation over which a spoken language model operates largely determines how much semantic competence survives.
Acoustic tokenizers such as SoundStream~\citep{soundstream} and EnCodec~\citep{encodec} use residual vector quantization for high-fidelity reconstruction, but their codes are optimized for signal fidelity rather than linguistic content, yielding token sequences that are long and only loosely related to text.
Semantic tokenizers derived from self-supervised models such as HuBERT~\citep{hubert} carry more linguistic information but discard the acoustic detail needed for expressive synthesis.
Hybrid designs attempt to span both.
SpeechTokenizer~\citep{speechtokenizer} distills semantic information into the first RVQ layer, CosyVoice~\citep{cosyvoice, cosyvoice2} supervises tokens with an ASR objective, and GLM-4-Voice~\citep{glm4-voice} derives an ultra-low-bitrate tokenizer from an ASR encoder.
All of these, however, produce token streams at a rate decoupled from the text tokenizer, leaving a length mismatch that the language model must absorb.

\subsubsection{Modality Integration Strategies}
Given this mismatch, systems address it in different ways.
Several mix text and speech tokens directly within a sequence. SpiritLM~\citep{spiritlm} interleaves at word level, Moshi~\citep{moshi} predicts time aligned text before audio through an inner monologue mechanism, Qwen2.5-Omni~\citep{qwen2-omni} routes the two through a Thinker-Talker split, and Mini-Omni ~\citep{mini-omni} employs parallel decoding streams for simultaneous text and audio emission, whereas Baichuan-Audio ~\citep{baichuan-audio} generates text and audio tokens in an interleaved manner.
Others instead absorb the resulting modality confusion with scale. GLM-4-Voice~\citep{glm4-voice} trains on 1 trillion tokens and Kimi-Audio~\citep{kimi-audio} trains on 13 million hours of audio.
A further group sidesteps the problem by constraining what gets retrained. Freeze-Omni~\citep{freeze-omni} keeps the LLM frozen and needs only 60K samples, and LLaMA-Omni~\citep{llama-omni, llama-omni2} attaches a streaming speech decoder to a frozen backbone, remaining competitive with GLM-4-Voice on far less speech data.

\subsubsection{Addressing Semantic Degradation}
Think Before You Talk~\citep{think-before-talk} diagnoses the failure directly.
Mixing speech tokens into the context dilutes the linguistic patterns acquired during text pretraining, so its TurnGuide method emits turn-level text guidance before speech to compensate, , eliminating modality-interleaving within the turn to preserve the precise time alignment essential for natural interactions..
TASTE~\citep{taste} addresses the mismatch at its source by using attention-based aggregation to construct one speech representation per text token.
Its word-level averaging for vocabulary alignment nonetheless discards fine grained acoustic detail and requires language specific segmentation rules, which limits multilingual scaling.
TASTE2 removes word-level averaging by sharing one text-token vocabulary across its components. Its modality aligned objective predicts one continuous audio latent for each text token, so it requires no additional interleaving along the token axis.

\subsection{Full-Duplex Dialogue Systems}
\label{ssec:rw_duplex}

Building a duplex system on top of such a model forces a second trade-off, orthogonal to the tokenization question.
End-to-end joint training lets information flow freely between listening and speaking but couples every capability into one set of weights, whereas pipelines keep components separately optimizable at the cost of the boundary losses noted above.
Recent surveys~\citep{fd-slm-survey, wavchat} organize the resulting design space along this axis, distinguishing engineered synchronization in modular systems from synchronization learned end-to-end.

\subsubsection{End-to-End Architectures}
Moshi~\citep{moshi} was the first to demonstrate real-time full-duplex dialogue, modeling user and system streams in parallel, but its tight coupling makes individual components hard to improve in isolation. SyncLLM~\citep{syncllm} gives a standard autoregressive LLM a sense of wall-clock time by interleaving fixed duration chunks of both speakers, predicting the user's chunk to mask network delay. OmniFlatten~\citep{omniflatten} flattens user and agent speech and text into a single sequence and progressively removes the text streams across training stages to cut latency. SALMONN-omni \citep{salmonn-omni} avoids discrete codecs entirely, operating on continuous embeddings so the model can listen to its own output for echo cancellation. NTPP \citep{ntpp} and SALM-Duplex \citep{salm-duplex} instead exploit dual channel recordings, the former through speaker independent Next Token Pair Prediction and the latter through channel fusion over a pretrained streaming encoder. Rather than paying for this coupled optimization with heavier speech domain training data, both systems are built to reduce that burden. SALMONN-omni matches or surpasses prior full duplex systems using substantially less training data, while SALM-Duplex skips speech pretraining altogether, markedly lowering its data requirement.

\subsubsection{Turn-Taking and Interruption}
Predicting when to speak has a long history independent of generative modeling.
TurnGPT~\citep{turngpt} predicts turn completion from lexical context, and Voice Activity Projection~\citep{vap} learns near-future voice activity of both speakers directly from audio, supporting real-time backchannel and turn-shift decisions.
Talking Turns~\citep{talking-turns} applies this lens to modern audio foundation models and finds them wanting, misjudging timing, interrupting aggressively, and rarely backchanneling, evidence that end-to-end training does not by itself confer competent turn management.
Full-Duplex-Bench~\citep{full-duplex-bench} turns this diagnosis into a standard measurement, scoring pause handling, backchanneling, and interruption directly, its v1.5 extension~\citep{full-duplex-bench-v15} adds overlap scenarios and prosodic adaptation, and FD-Bench~\citep{fd-bench} targets robustness under frequent disruption.
TASTE2 VoiceBot takes the modular route. Because TASTE2 aligns audio latents with text tokens, components can exchange both forms of information and make turn boundary decisions at the text-token level. The deployed system processes user speech incrementally, begins ordinary responses after detected speech end, and supports user barge-in during playback.

\subsection{Paralinguistic Control and Adaptation}
\label{ssec:rw_para}

\subsubsection{Paralinguistic Control}
One line of work gives the model an explicit style command and measures whether it complies.
GLM-4-Voice~\citep{glm4-voice} exposes emotion, intonation, and speaking rate to natural-language instruction, so a user can ask the agent to slow down or to sound cheerful and the model only has to execute the request.
VStyle~\citep{vstyle} issues the instruction as speech rather than text and reports that current models face clear limitations in controllable style adaptation.
Three of its four categories, acoustic attributes, natural-language instruction, and role play, state the target explicitly, while the fourth asks for implicit empathy and so already crosses into inference from unstated cues.
SpeechParaling-Bench~\citep{speechparaling-bench} widens the coverage further, testing over 100 paralinguistic features across fine-grained control, intra-utterance variation, and context aware adaptation, and finds that even frontier models fall short of consistent control.
Style following of this kind establishes that a model can render a delivery on request. It says little about whether the model can choose one when the user names nothing.

\subsubsection{Paralinguistic Adaptation}
A second line removes the command, so the model must infer an appropriate style from the paralinguistic cues carried in the user's speech, including emotion but also prosody, speaking rate, and speaker attributes.
ProsodyLM~\citep{prosodylm} shows that this perception can emerge from pretraining alone on word-level prosody tokens, with no instruction directing the model toward emotion or contrastive focus.
TELEVAL~\citep{televal} argues that this is the capability that matters most for a conversational agent, extracting cues from user speech and responding appropriately without being told to.
Adaptation also affects response content and not only delivery, since a user who sounds distressed warrants different words as well as a gentler tone.
EMO-Reasoning~\citep{emo-reasoning} covers the emotional case across turns, scoring the coherence of emotional transitions in extended dialogue, while ParaS2S~\citep{paras2s} scores the broader paralinguistic fit within a turn, judging whether a response matches its input in content and style together and optimizing that fit at the waveform level with reinforcement learning. EchoMind \citep{echomind} isolates the content versus delivery distinction directly. Its tasks hold transcripts fixed and vary only vocal style, showing that models can perceive a paralinguistic cue without translating it into an appropriately adapted response.

The empirical scope of this report is explicit paralinguistic control after dialogue fine-tuning, complemented by a pre-fine-tuning probe of whether a Spoken LM can continue a feature already present in a speech opening. We do not evaluate implicit paralinguistic adaptation, which remains future work.

\section{TASTE2 Model and Training}
\label{sec:model_and_training}

TASTE2 builds on TASTE~\citep{taste} and modifies three aspects of its architecture for incremental dialogue: word-level averaging, vocabularies across components, and utterance-level synthesis (§\ref{subsubsec:revisiting_taste}).

TASTE2 introduces an aligned text-token space (§\ref{subsubsec:aligned_token_space}) and incremental synthesis through CosyVoice2 (§\ref{subsubsec:streaming_output}). Dialogue fine-tuning then teaches the TASTE2 Spoken LM turn-boundary and interruption patterns (§\ref{subsubsec:dialogue_finetuning}).

\begin{figure}[!ht]
\centering
\includegraphics[width=\columnwidth]{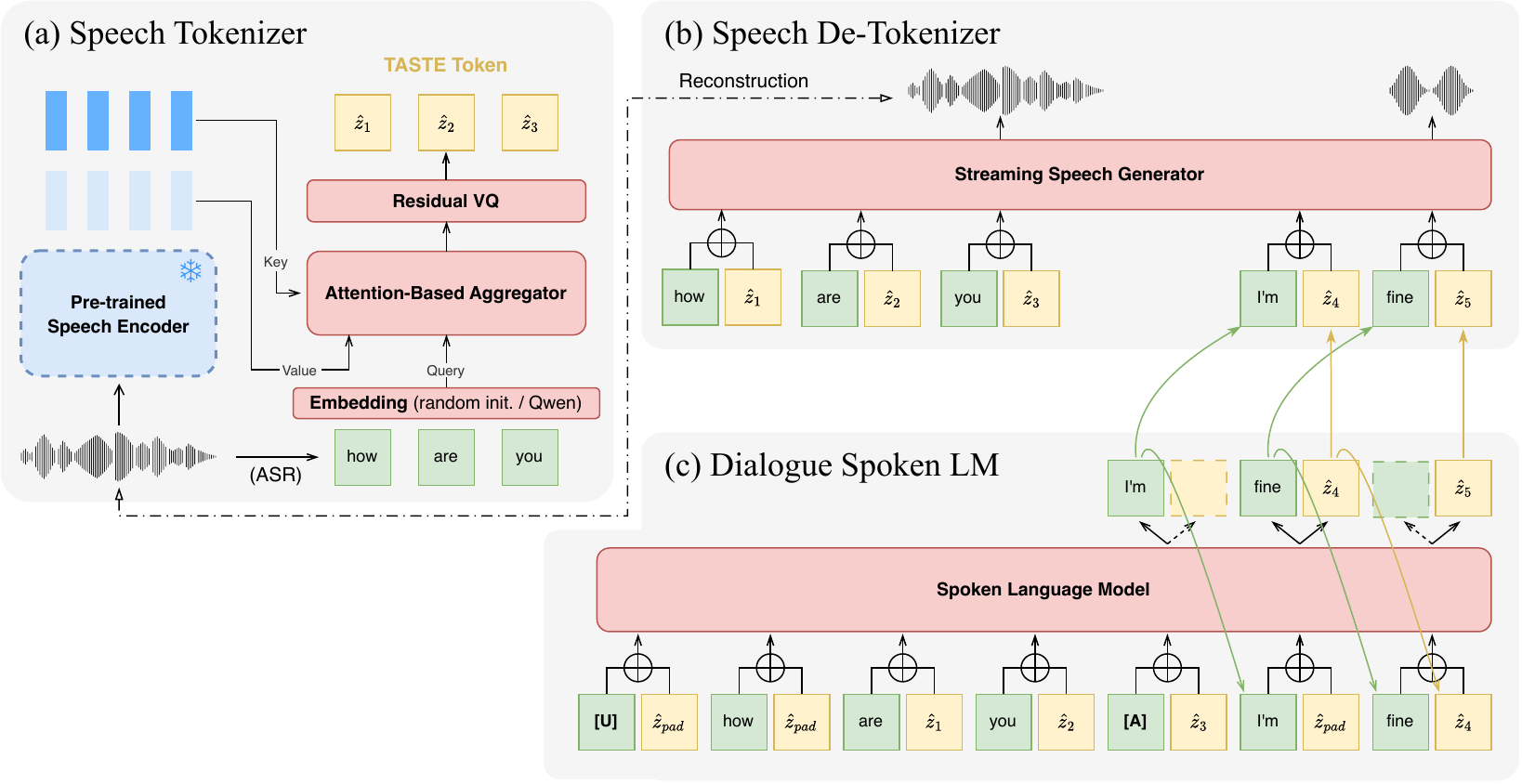}
\caption{TASTE2 architecture. The Speech Tokenizer uses a frozen Distil-Whisper encoder to extract continuous audio latents aligned with text tokens. The TASTE2 Spoken LM autoregressively predicts text-token--audio-latent pairs. The Speech Detokenizer maps each pair to S3 units, which the frozen CosyVoice2 flow-matching model and vocoder synthesize incrementally. The Attention-Based Aggregator initializes from Whisper's decoder, the Spoken LM from Qwen2.5-7B, and the Speech Detokenizer's flow-matching model from Qwen2.5-0.5B; the Residual VQ is randomly initialized.}
\label{fig:taste2_architecture}
\end{figure}

Figure~\ref{fig:taste2_architecture} shows the three model components. The Speech Tokenizer extracts one continuous audio latent per text token. The TASTE2 Spoken LM autoregressively predicts aligned text-token--audio-latent pairs. The Speech Detokenizer maps those pairs to S3 units, which the frozen CosyVoice2 flow-matching model and vocoder convert to audio. Training proceeds in three stages: representation learning, Spoken LM pretraining, and dialogue fine-tuning.

\subsection{Model Architecture}
\label{subsec:model_arch}

\subsubsection{Original TASTE Architecture}
\label{subsubsec:revisiting_taste}

TASTE~\citep{taste} uses text-aligned speech tokenization for joint text-speech spoken language modeling. The framework comprises three components: a Speech Tokenizer consisting of an encoder, attention-based aggregator, and Residual Vector Quantizer; a Speech Detokenizer composed of an S3 unit decoder and vocoder; and a Text-Aligned Spoken Language Model (TASLM). An end-to-end cross-attention mechanism establishes a one-to-one correspondence between text and speech tokens at the token level. It uses soft alignment cues from the ASR encoder hidden states to aggregate multiple audio frames into text-aligned speech tokens. During training, word-level averaging connects the Speech Tokenizer and Speech Detokenizer vocabularies to the downstream Spoken LM vocabulary. TASLM uses a pretrained LLaMA text language model as its backbone and applies Low-Rank Adaptation. It takes aligned text-token sequences as input and autoregressively generates text-speech token pairs.

TASTE performs averaging over word boundaries, so its Speech Tokenizer and Spoken LM operate with different text-token boundaries, and dialogue templates and token-level turn-boundary decisions require conversion between them. The original synthesis stack, based on CosyVoice~\citep{cosyvoice}, produces audio only after receiving a complete utterance. TASTE2 changes the vocabulary, token alignment, and synthesis timing described below.

\subsubsection{Aligned Token Space}
\label{subsubsec:aligned_token_space}

TASTE uses different vocabularies in its Whisper-based Speech Tokenizer and LLaMA backbone, with word-level averaging between the two sequences.

TASTE2 instead uses the Spoken LM tokenizer throughout the model. The Speech Tokenizer, TASTE2 Spoken LM, and Speech Detokenizer therefore share the same text-token vocabulary and no longer require word-level averaging. Our implementation uses a Qwen2.5-7B family backbone for the Spoken LM (base or instruction-tuned, depending on the instruction-following initialization strategy; \S\ref{subsubsec:if_strategies}) and adapts the Qwen2.5-0.5B text-to-S3 architecture from CosyVoice2~\citep{cosyvoice2} as the Speech Detokenizer.

Our implementation modifies Distil-Whisper's architecture. The encoder remains frozen and retains its pretrained weights. The decoder's input embedding layer is replaced with the Spoken LM tokenization vocabulary and randomly initialized. The language modeling head is removed, and the decoder outputs continuous embeddings directly to the Residual Vector Quantizer (RVQ). This training configuration does not add a separate initialization scheme or warm-up strategy.

The Speech Tokenizer, Spoken LM, and Speech Detokenizer use the same LM tokenizer, and each text token corresponds to one audio latent. Each position in a standard text dialogue template therefore contains a text token and its aligned acoustic representation, so no additional interleaving along the token axis is required.

\subsubsection{Support for Streaming Output}
\label{subsubsec:streaming_output}

Incremental dialogue requires synthesis to begin before the complete response is available. TASTE2 reuses the CosyVoice2 streaming pipeline~\citep{cosyvoice2}, which contains a text-to-S3 model produces S3 units at 25~Hz, a chunk-aware causal flow-matching model produces mel spectrograms at 50~Hz, and a vocoder produces audio waveforms.

We only train the Speech Detokenizer and make it learn to autoregressively map aligned text-token and audio-latent pairs from the TASTE2 Spoken LM into the CosyVoice2 S3 space. To enable streaming, it processes inputs in chunks by producing 15 S3 units for every 5 input pairs, allowing audio to be synthesized without waiting for the full utterance. The downstream flow-matching model and vocoder are inherited from CosyVoice2 and remain frozen.

\subsection{Data Processing}
\label{subsec:data_processing}

\subsubsection{Data Collection}
\label{subsubsec:data_collection}

The three training stages use different data. Stages~1 and 2 use large speech corpora for representation learning and Spoken LM pretraining. Stage~SFT uses conversational data containing turn transitions, interruptions, and paralinguistic instructions.

Following TASTE's data collection strategy, we use two datasets for pretraining. The \textbf{Emilia dataset}~\citep{emilia} provides 40,000 hours of English in-the-wild speech with pseudo-labeled transcriptions and includes background noise, reverberation, and multiple-speaker conditions. The \textbf{LibriTTS dataset}~\citep{libritts} provides 600 hours of reading-style speech with human-annotated transcriptions. Combined, these datasets total approximately 40,600 hours of training data.

Stage SFT dialogue fine-tuning uses two data sources. The \textbf{DeepDialogue dataset} comprises 37,000 multi-turn dialogues (223,673 turns, totaling 466 hours) spanning 41 domains (e.g., customer service, technical support, and casual conversation) with 20 emotion labels (e.g., happy, frustrated, and empathetic). These dialogues were generated by an ensemble of nine large language models, followed by human annotation and quality filtering. Each dialogue was synthesized with a TTS system using the associated emotion label.

We also generated 106,507 \textbf{synthetic dialogues} (557,770 turns, totaling 1,027 hours) in two categories: general conversation data for full-duplex interaction modeling (48,000 dialogues with 388,021 turns over 778 hours) and paralinguistic control data (58,507 dialogues with 169,749 turns over 249 hours). The former includes interruption scenarios, and the latter includes explicit paralinguistic control annotations. The generation pipeline is detailed in the following subsection.

Table~\ref{tab:data_summary} summarizes the training data. Stages~1 and 2 use 40,600 hours of speech data. Stage~SFT uses 143,507 dialogues comprising 781,443 turns and 1,493 hours of speech: 466 hours from DeepDialogue and 1,027 hours from synthetic dialogue data. The Stage~SFT data include turn-taking, interruption, and paralinguistic-control annotations.

\begin{table}[!ht]
\centering
\caption{Training data used in Stages 1--2 and SFT. Dialogue and turn counts apply only to the conversational corpora used for SFT.}
\label{tab:data_summary}
\begin{tabular}{llrrr}
\toprule
\textbf{Stage} & \textbf{Dataset} & \textbf{Dialogues} & \textbf{Turns} & \textbf{Duration (h)} \\
\midrule
\multirow{3}{*}{Stage 1 \& 2}
    & Emilia & -- & -- & 40,000 \\
    & LibriTTS & -- & -- & 600 \\
    \cmidrule(lr){2-5}
    & \textbf{Subtotal} & \textbf{--} & \textbf{--} & \textbf{40,600} \\
\midrule
\multirow{4}{*}{Stage SFT}
    & DeepDialogue & 37,000 & 223,673 & 466 \\
    & General Conversation & 48,000 & 388,021 & 778 \\
    & Paralinguistic Control & 58,507 & 169,749 & 249 \\
    \cmidrule(lr){2-5}
    & \textbf{Subtotal} & \textbf{143,507} & \textbf{781,443} & \textbf{1,493} \\
\bottomrule
\end{tabular}
\end{table}

\subsubsection{Synthetic Data Generation Strategy}
\label{subsubsec:synthetic_data}

Naturally occurring dialogue corpora rarely provide explicit interruption boundaries and controlled paralinguistic annotations. We therefore supplement DeepDialogue with 1,027 hours of synthetic data: 778 hours of general conversations with turn transitions and interruptions, and 249 hours of conversations with explicit paralinguistic controls.

\textbf{General Conversation Data.} We use a six-step pipeline to synthesize this data. (1) generates scenarios with GPT-4o model, (2) produces five 6--10 turns dialogues per scenario with Llama-3.3-70B, (3) uses GPT-4o to rewrite a subset with user interruptions, (4) inserts hesitations, fillers, and false starts, (5) filters dialogues with a GPT-4o judge, and (6) synthesizes the retained dialogues with CosyVoice2 to create the final audio data. We draw from 50 public reference speakers and use CosyVoice2's instruction-following mode to vary the output style.
The resulting data contain interruptions at semantic boundaries, topic changes during Assistant responses, spontaneous speech disfluencies, and different emotions for User and Assistant turns.

\textbf{Paralinguistic Control Data.} We use a four-step pipeline to synthesize this data (1) generates scenarios with GPT-4o model, (2) alternates Llama-3.3-70B user turns with GPT-4o assistant turns to produce five approximately 10 turns dialogues per scenario, (3) filters the dialogues with a GPT-4o judge, and (4) synthesizes them with IndexTTS. In this data, the user requests a change in speaking style and the assistant continues the conversation using the requested emotion in randomly selected turns. Synthesis also uses the same pool of 50 reference speakers as the general conversation data and IndexTTS's emotion-vector mode to enforce the annotated emotion.

For the step using GPT-4o to filter both data pipeline, GPT-4o scores each dialogue from 1 to 5 on six criteria: role and scenario adherence, language naturalness and variety, emotion and speed tag placement, formatting, coherence of inserted overlaps and disfluencies, and absence of extraneous stage directions. We sum the six ratings and retain the three highest-scoring dialogues for each scenario.

\subsection{Training Methodology}
\label{subsec:training_methodology}

TASTE2 training proceeds through three stages. \textbf{Stage~1} jointly trains the Speech Tokenizer and Speech Detokenizer to establish the audio-latent space. \textbf{Stage~2} adapts a pretrained text language model into the TASTE2 Spoken LM with LoRA, training it to predict aligned text-token--audio-latent pairs. \textbf{Stage~SFT} fine-tunes the Spoken LM on dialogue data (\S\ref{subsubsec:dialogue_finetuning}).

The two training configurations differ in how instruction-following (IF) capability is introduced (\S\ref{subsubsec:if_strategies}). \textbf{Approach 1 (Base $\rightarrow$ Merge $\rightarrow$ SFT)} starts Stage~2 from Qwen2.5-7B and transfers IF capability through model merging before Stage~SFT. \textbf{Approach 2 (Instruct $\rightarrow$ SFT)} starts Stage~2 directly from Qwen2.5-7B-Instruct and does not perform model merging.

\subsubsection{Spoken Language Model Pretraining}
\label{subsubsec:slm_pretraining}

Stages 1 and 2 follow the TASTE training procedure. Stage~1 jointly trains the Speech Tokenizer to extract speech representations from audio and the Speech Detokenizer to reconstruct audio from paired text tokens and speech representations (Figure~\ref{fig:taste2_architecture}). In Stage~2, the language model is initialized from Qwen2.5-7B for Approach~1 and Qwen2.5-7B-Instruct for Approach~2. Both configurations use the same LoRA-based fine-tuning objective to generate aligned text tokens and audio latents (Figure~\ref{fig:taste2_architecture}).

\textbf{Residual Vector Quantization.} During Spoken LM training, the audio-latent loss uses the final latent synthesized by the Residual Vector Quantizer (RVQ) as its target rather than a sequence of quantized indices. The RVQ is positioned between the Speech Tokenizer decoder and the audio-latent target. Section~\ref{subsec:ablation_rvq} reports the comparison between configurations with and without RVQ.

\textbf{Modality-aligned training strategy.} TASTE2 maintains text and audio as paired channels at the text-token rate. Each input position combines a text-token embedding with its aligned audio-latent embedding through a weighted sum. The latent is not inserted as an additional sequence element. For autoregressive generation, the audio target is shifted by one position so that the text token is predicted before its corresponding audio latent. Text tokens and audio latents serve as the prediction targets for content and acoustic realization, respectively.

\subsubsection{Instruction-Following Initialization Strategies}
\label{subsubsec:if_strategies}

Approach~1 introduces IF capability after Stage~2 by transferring an instruction-following task vector to the pretrained Spoken LM backbone. This task-vector arithmetic requires no additional training data or compute at the merging step. The merged model then serves as the initialization checkpoint for Stage~SFT.

\textbf{Task-vector merging.} We extract the language model backbone from the pretrained Spoken LM checkpoint by isolating its language model weight tensors from the full Spoken LM state dict. Let $W_{\text{SLM}}$ denote these extracted backbone weights, $W_{\text{base}}$ the weights of Qwen2.5-7B, and $W_{\text{instruct}}$ those of Qwen2.5-7B-Instruct. We construct a sparsified IF task vector using TIES merging~\citep{ties}: the raw task vector $\tau = W_{\text{instruct}} - W_{\text{base}}$ is sparsified by retaining only the top $p$ parameters by absolute magnitude and zeroing out the rest,
$$\hat{\tau} = \mathrm{TopK}_{p}(\tau), \quad p = 0.3$$
and the merged backbone is computed as:
$$W_{\text{merged}} = W_{\text{SLM}} + \lambda \cdot \hat{\tau}, \quad \lambda = 0.5$$
The merged weights are injected back into the Spoken LM checkpoint structure, replacing the original backbone weights. This merged checkpoint is then used as the starting point for Stage~SFT (\S\ref{subsubsec:dialogue_finetuning}). In Approach~2, the Spoken LM is initialized from Qwen2.5-7B-Instruct at Stage~2 and proceeds to the same Stage~SFT directly, without task-vector arithmetic.

\subsubsection{Dialogue Supervised Fine-Tuning}
\label{subsubsec:dialogue_finetuning}

Apart from their Stage~2 initialization and the merging step, both approaches use the same speec learning objectives, dialogue data, and Stage~SFT procedure. Stage~SFT extends single-utterance generation to multi-turn dialogue, using a structured template and the existing autoregressive objective to train response generation, turn-boundary estimation, and how to appropriately react to interruption sequences.

\begin{figure}[!ht]
\centering
\includegraphics[width=\columnwidth]{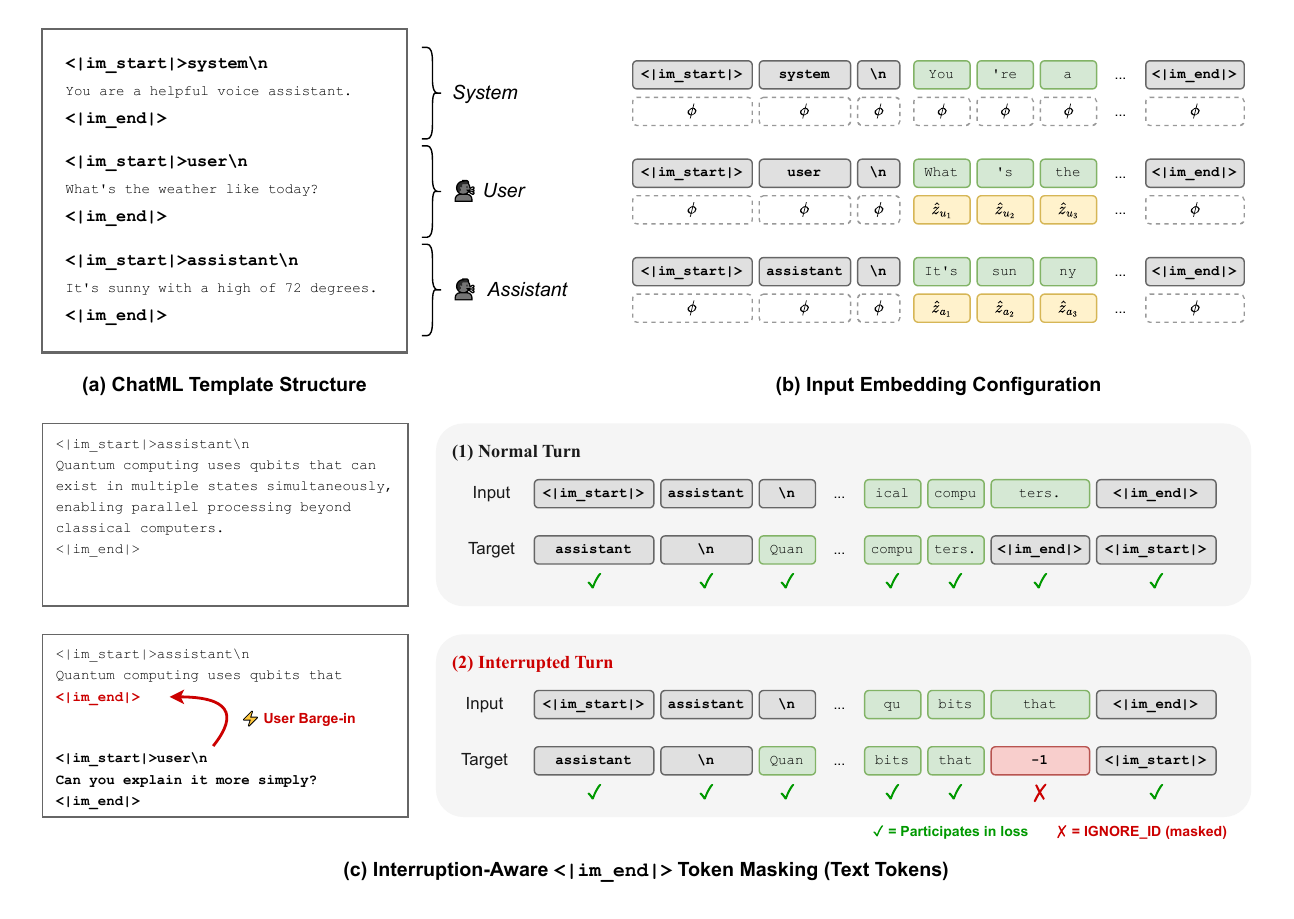}
\caption{Dialogue template and interruption-aware training strategy. \textbf{(a)} ChatML segments for System, User, and Assistant, delimited by \texttt{<|im\_start|>} and \texttt{<|im\_end|>}. \textbf{(b)} One-to-one pairing of text tokens (green) and audio latents (yellow); special tokens use zero-filled audio embeddings (white, dashed). \textbf{(c)} Complete turns compute loss over all tokens, whereas interrupted Assistant turns set the loss weight of the terminal \texttt{<|im\_end|>} token to 0 (marked $\times$).}
\label{fig:dialogue_template}
\end{figure}

\textbf{Template Design and Input Configuration.} Figure~\ref{fig:dialogue_template} illustrates our dialogue template
design, input embedding configuration, and interruption-aware training
strategy. We adopt the same ChatML-style dialogue template used natively by the Qwen2.5 tokenizer, whose special tokens \texttt{<|im\_start|>} and \texttt{<|im\_end|>} delimit conversational segments. Audio does not introduce a separate special token but instead attaches to each text-token position as an aligned audio latent. Each dialogue comprises three segment types: \emph{System} (instruction prompt), \emph{User} (user utterances), and \emph{Assistant} (model responses). Segments are arranged sequentially without intervening newlines, with each segment following the pattern \texttt{<|im\_start|>}\textit{role}\texttt{$\backslash$n}\textit{content}\texttt{<|im\_end|>} where \textit{role} $\in$ \{\texttt{system}, \texttt{user}, \texttt{assistant}\}. During training, input embedding configuration varies by segment type: \emph{System segments} contain text tokens with zero-filled audio latents because system prompts are non-voiced instructional content; \emph{User segments} pair text tokens from ASR transcription with aligned audio latents extracted by the TASTE2 Speech Tokenizer; and \emph{Assistant segments} pair text tokens with audio latents from ground-truth responses. The training objective applies unified next-token prediction and predicts both the next text token and its aligned audio latent.

\textbf{Supervised Fine-Tuning Paradigm.} Stage~2 and Stage~SFT share the same unified autoregressive objective over the template above. The loss is computed over User and Assistant segments; System segments provide conditioning context and are excluded from the loss. Interruption examples receive special loss treatment, and the resulting \texttt{<|im\_end|>} distribution drives the turn-decision gate. Both are described below.

\textbf{Interruption-Aware Loss Design.} Full-duplex dialogue training contains both normal turn-taking and interruption examples. As shown in Figure~\ref{fig:dialogue_template}, in normal examples, the Assistant's response concludes with the \texttt{<|im\_end|>} token, and loss is computed over all tokens. In interruption examples, the Assistant's utterance is truncated at the User's barge-in position. For these truncated segments, the loss weight of the terminal \texttt{<|im\_end|>} token is set to 0, while all preceding content tokens retain their loss. At inference time, this design allows generation to stop immediately upon detecting the onset of user speech.

\textbf{Model-Gated Turn Decisions.} In the training data, a completed User segment is followed by \texttt{<|im\_end|>}, whereas an incomplete one continues with another content token. The model therefore learns to place probability mass on \texttt{<|im\_end|>} exactly where a user turn ends, and that probability is the signal consumed by the deployed turn-taking gate (\S\ref{subsubsec:turn_taking}).

\textbf{Training Configuration.} During Stage~SFT, the Speech Tokenizer, Speech Detokenizer, CosyVoice2 flow-matching model, and vocoder remain frozen. The Spoken LM continues from the Stage~2 checkpoint and is fine-tuned with LoRA. Training data comprise three categories: normal multi-turn dialogues from DeepDialogue and synthetic general conversation data, synthetic interruption scenarios, and dialogues from the synthetic paralinguistic control data (\S\ref{subsubsec:data_collection}, Table~\ref{tab:data_summary}).

\section{TASTE2 VoiceBot: Incremental Deployment Architecture}
\label{sec:system_design}

TASTE2 VoiceBot deploys the model from §\ref{sec:model_and_training} as an incremental pipeline. It processes user speech as it arrives, decides when to take the floor, streams its response, and stops when the user barges in. This section describes the services that realize these behaviors and the deployment decisions they require.

\subsection{Overall System Architecture}
\label{subsec:system_architecture}

\begin{figure}[!ht]
\centering
\includegraphics[width=\columnwidth]{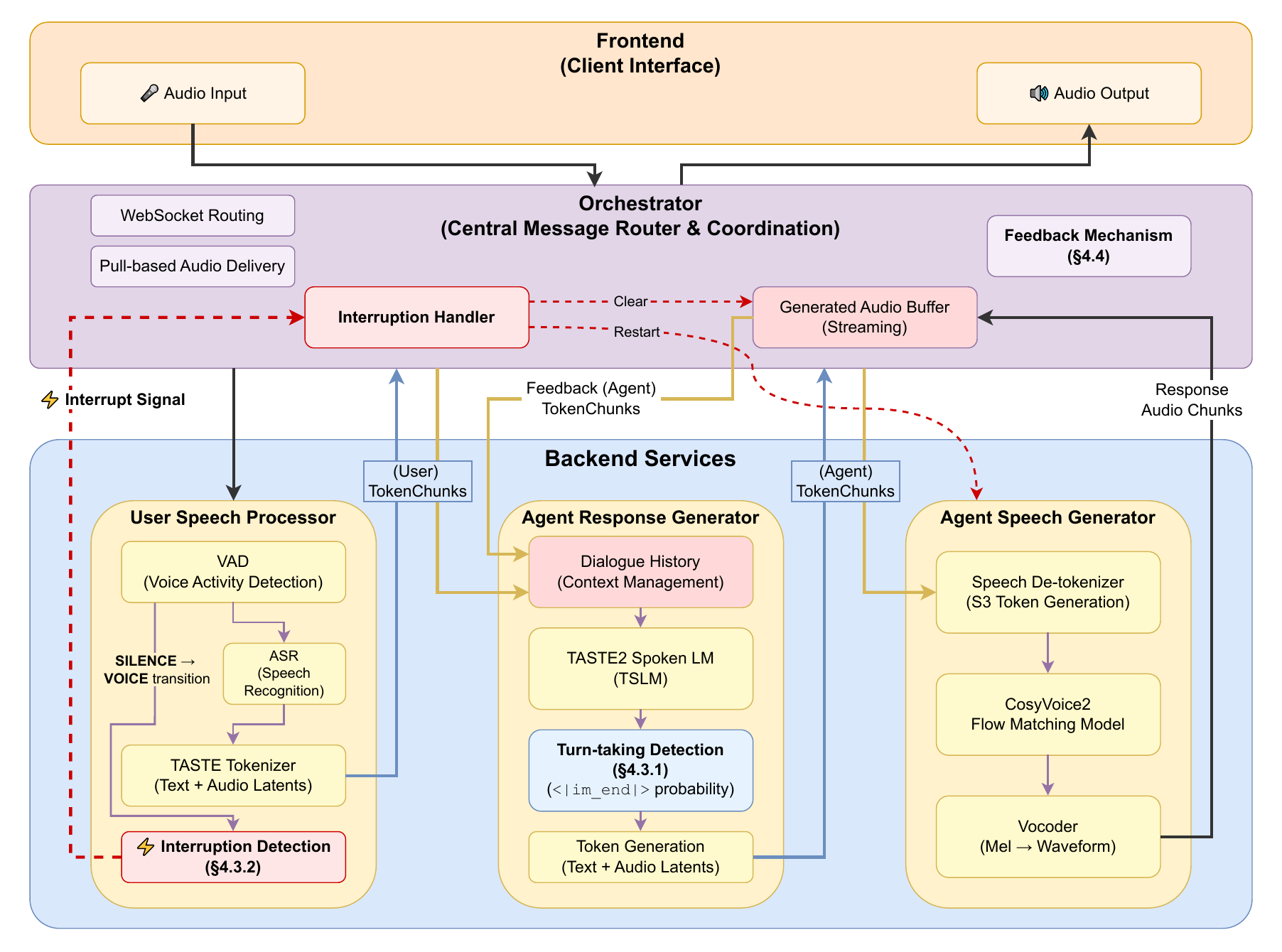}
\caption{TASTE2 VoiceBot system architecture. Four WebSocket services form the dialogue pipeline: the \textbf{User Speech Processor} performs VAD, ASR, TASTE tokenization, and interruption detection (§\ref{subsubsec:interruption_handling}); the \textbf{Agent Response Generator} maintains dialogue context and applies model-gated turn decisions (§\ref{subsubsec:turn_taking}); the \textbf{Agent Speech Generator} streams detokenized and vocoded speech; and the \textbf{Orchestrator} coordinates delivery, interruption handling, and playback-synchronized feedback (§\ref{subsec:audio_playback_feedback}).}
\label{fig:system_architecture}
\end{figure}

TASTE2 VoiceBot employs a WebSocket-based microservice architecture comprising four main services (illustrated in Figure~\ref{fig:system_architecture}). Each service handles a distinct stage of the conversation pipeline, communicating asynchronously through WebSocket connections.

The \textbf{User Speech Processor} converts raw audio into aligned text tokens and audio latents. Voice activity detection (VAD) identifies speech and silence, automatic speech recognition (ASR) provides the transcription, and the TASTE2 Speech Tokenizer extracts the audio latents. A VAD transition from silence to voice during system playback emits a barge-in event before ASR completes.

The \textbf{Agent Response Generator} maintains dialogue context and runs the TASTE2 Spoken LM. It receives \texttt{TokenChunk} messages containing aligned text tokens and audio latents. After a VAD speech-end signal, it uses the probability of \texttt{<|im\_end|>} to decide whether the user has completed a semantic turn (§\ref{subsubsec:turn_taking}); only then does it autoregressively predict response pairs. Feedback from the Orchestrator adds only delivered response pairs to the dialogue history.

The \textbf{Agent Speech Generator} maps response pairs to S3 units with the Speech Detokenizer, then uses the CosyVoice2 flow-matching model and vocoder to produce audio. It synthesizes complete input chunks incrementally and stops when it receives an interruption event (§\ref{subsec:streaming_pipeline}).

\textbf{Orchestrator} coordinates all services and manages client interactions. The service acts as a central message router, orchestrating communication between User Speech Processor, Agent Response Generator, and Agent Speech Generator. Crucially, the Orchestrator implements pull-based audio delivery and synchronized feedback mechanism (§\ref{subsec:audio_playback_feedback}), buffering audio from Agent Speech Generator, delivering chunks only upon client request, and feeding back corresponding agent tokens to Agent Response Generator only after delivery. This design ensures dialogue context remains synchronized with the user's actual auditory experience, even during interruptions.

\subsection{Streaming Processing Pipeline}
\label{subsec:streaming_pipeline}

Full-duplex voice conversation is sensitive to accumulated response latency. Traditional pipelines process ASR, language model inference, and speech synthesis sequentially, with each stage waiting for the previous one to complete. TASTE2 therefore adopts a latency-oriented incremental pipeline.

TASTE2 breaks each pipeline stage's work into small token chunks and passes results between stages through queues, so a downstream stage can begin processing as soon as an upstream chunk is ready instead of waiting for a complete utterance. TASTE2 addresses this challenge through a three-service concurrent streaming architecture. The pipeline comprises three services: (1) \textbf{User Speech Processor}: performs VAD, ASR, and TASTE tokenization, processing audio chunks incrementally and producing partial token representations; (2) \textbf{Agent Response Generator}: receives streaming user token chunks. Although the architecture can consume token chunks as they arrive, the deployed configuration invokes the turn-decision gate only when VAD detects speech end (voice → silence transition). The Spoken LM then either starts an Assistant segment or remains silent, avoiding repeated generation attempts while the user speaks; and (3) \textbf{Agent Speech Generator}: uses TASTE2's streaming output synthesis (§\ref{subsubsec:streaming_output}) to convert agent tokens into S3 units and then produce audio waveforms incrementally through CosyVoice2's chunk-aware flow matching model and vocoder. The services operate independently and pass partial results downstream as they become available.

This streaming capability is realized by running each stage of the User Speech Processor concurrently rather than sequentially. VAD continuously scans incoming 100ms audio chunks and a streaming ASR component asynchronously transcribes accumulated audio as it arrives, decoupled from voice activity detection through an intermediate queue. Then, TASTE tokenization consumes the streaming ASR output through a second queue to produce aligned text tokens and audio latents. This queue-based communication prevents blocking between stages and each stage processes immediately when input arrives, without waiting for downstream consumption. The Agent Speech Generator applies the same concurrent design. One component continuously receives S3 units from the Spoken LM, decoupling it from generation, while a second component executes GPU audio synthesis. This design ensures GPU operations do not block the pipeline, maintaining streaming characteristics throughout.

TASTE2 VoiceBot adapts the originally utterance level Speech Tokenizer for incremental operation. Its causal decoder predicts the audio latent at position $i$ from only the first $i$ text tokens and the corresponding audio. VAD groups 100ms audio chunks into 600ms batches, after which the Speech Tokenizer extracts aligned text tokens and audio latents and sends them downstream. TASTE2 VoiceBot can therefore accumulate dialogue context while the user speaks, although ordinary response generation remains gated by detected speech end.

\subsection{Turn and Interruption Mechanisms}
\label{subsec:full_duplex_capabilities}

\subsubsection{VAD Triggered and Model-Gated Turn Decisions}
\label{subsubsec:turn_taking}

TASTE2 VoiceBot separates the acoustic trigger from the semantic turn decision. A VAD voice-to-silence transition (100~ms of silence) triggers evaluation, but does not by itself start an Assistant response. The Agent Response Generator then applies the turn-boundary distribution learned by the TASTE2 Spoken LM as a semantic gate, without adding a separate turn-taking classifier.

During training (§\ref{subsubsec:dialogue_finetuning}), the model learns implicit turn-end probability patterns through exposure to User segments. When a user finishes speaking, the dialogue sequence transitions to \texttt{<|im\_end|>} followed by \texttt{<|im\_start|>assistant} and the model learns to assign high probability to \texttt{<|im\_end|>} after final token of the sentence. Conversely, when a user is in the  middle of the  utterance, the model assigns low probability to \texttt{<|im\_end|>} and high probability to other continuation tokens. Through this pattern, \texttt{<|im\_end|>} probability implicitly encodes turn-end likelihood.

After the VAD trigger, the Agent Response Generator evaluates \texttt{<|im\_end|>} after the accumulated user tokens. The system supports two gate conditions: (1) \textbf{Top-k detection}: checking whether \texttt{<|im\_end|>} appears in the top-$k$ predictions ($k=800$); and (2) \textbf{Probability threshold}: checking whether \texttt{<|im\_end|>} probability exceeds a threshold $b$ ($b=0.3$). If either condition identifies a turn end, the model enters the Assistant segment; otherwise, it remains silent and awaits more user speech. Because the gate is evaluated only after VAD detects speech end, the deployed configuration never speaks during user speech.

Punctuation marks also pose challenges for turn-taking detection. They may signal sentence endings or mid-sentence pauses. TASTE2 employs a lookahead mechanism to resolve this ambiguity. When the predicted token is punctuation, the system uses KV cache to predict the next token and applies \texttt{<|im\_end|>} detection to the second token. This lookahead enables the system to distinguish sentence-final punctuation (e.g., `.` followed by high \texttt{<|im\_end|>} probability) from mid-sentence punctuation (e.g., `,` followed by low \texttt{<|im\_end|>} probability), improving turn-end detection accuracy in natural conversation.

\subsubsection{Interruption Handling}
\label{subsubsec:interruption_handling}

TASTE2 VoiceBot handles user barge-in through early VAD detection and coordinated cleanup. When the User Speech Processor's VAD detects the speaker voice (silence → voice transition) during agent speech, the system immediately emits an interruption event. This early signal (before ASR processing) enables rapid response to barge-in, minimizing latency. 
Once interruption is detected, the Orchestrator executes implement a four step system cleanup flow. (1) \textbf{Close Agent Speech Generator WebSocket}: immediately halting audio generation, preventing further resource consumption; (2) \textbf{Clear Audio Buffer}: discarding all unplayed audio chunks, preventing stale content delivery to the client; (3) \textbf{Reset interruption event flag}: preparing the event state for the next potential interruption; and (4) \textbf{Reconnect Agent Speech Generator}: establishing a new WebSocket connection for the next turn. This cleanup flow ensures the system rapidly recovers from interruption, ready to process the user's new input.

\subsection{Audio Playback and Feedback Mechanism}
\label{subsec:audio_playback_feedback}

TASTE2 VoiceBot must keep its dialogue history consistent with audio playback. Dual-channel models such as Moshi~\citep{moshi}, NTPP~\citep{ntpp}, and SALM-Duplex~\citep{salm-duplex} represent user and agent speech separately. TASTE2 VoiceBot instead tracks a single sequence of aligned text tokens and audio latents, while synthesis and playback proceed asynchronously.

Text-token--audio-latent pairs do not map individually to waveform segments of a fixed duration. Instead, the playback audio is synthesized at a constant 25 Hz frame rate for groups of input pairs, meaning that playback boundaries do not need to align strictly with individual text tokens. TASTE2 VoiceBot assigns a minimum number of input pairs to each audio chunk, establishing a coarse correspondence that can be tracked during delivery.

This architectural characteristic creates practical synchronization problems. Speech generation speed typically exceeds playback rate, enabling the system to rapidly produce complete responses. However, users may interrupt before playback completes. If all generated tokens are immediately fed back to dialogue context, the system's internal state will contain agent utterances the user never heard, causing the agent to later reference content the user never actually received and breaking conversational coherence.

\begin{figure}[!ht]
\centering
\includegraphics[width=0.85\columnwidth]{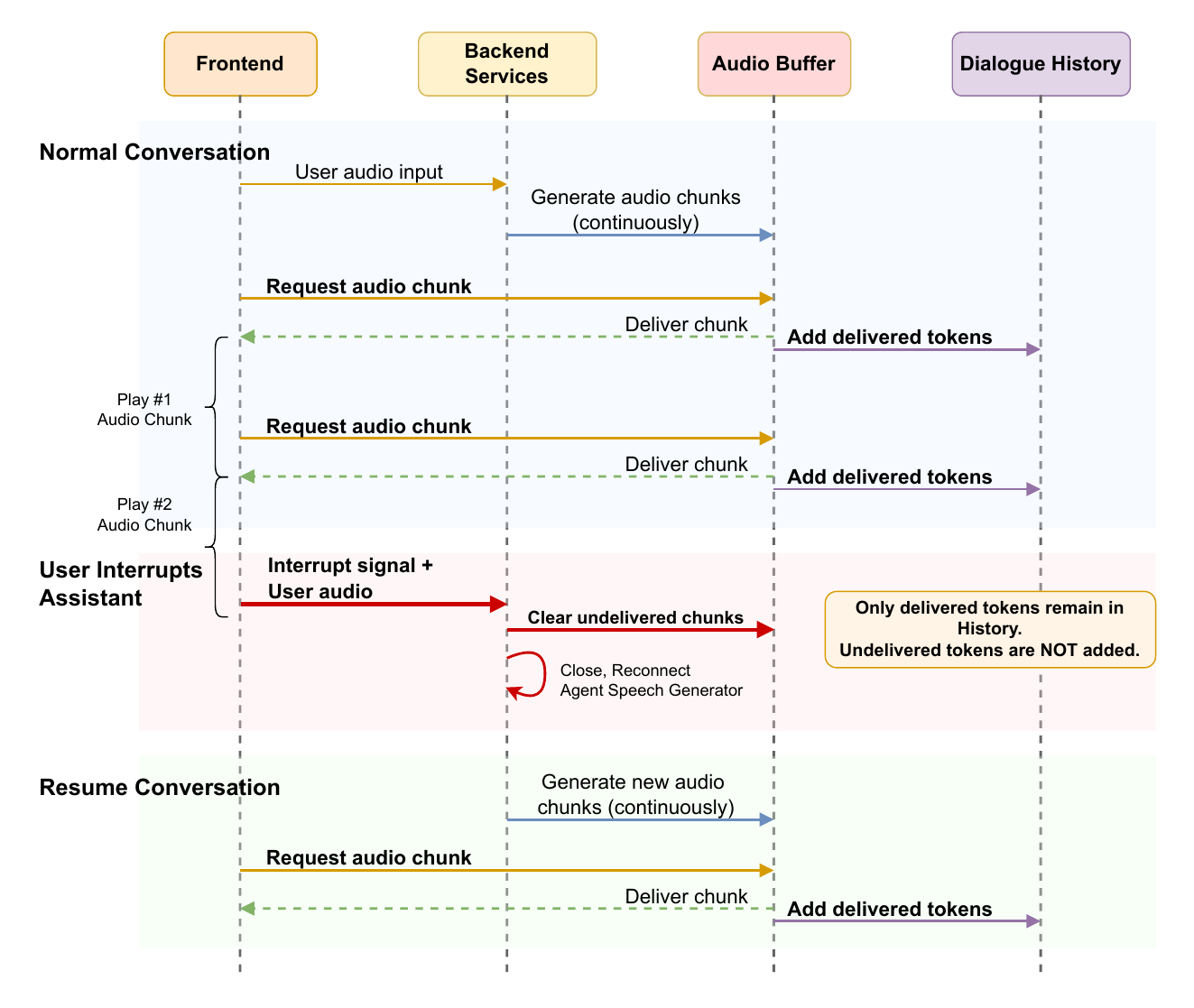}
\caption{Pull-based audio delivery and synchronized feedback. The frontend requests chunks from the Audio Buffer, and the corresponding tokens enter Dialogue History only after delivery. Upon user interruption, the Orchestrator clears undelivered chunks and closes the Agent Speech Generator connection; tokens for those chunks are excluded from the dialogue context.}
\label{fig:audio_feedback_mechanism}
\end{figure}

TASTE2 addresses this challenge through pull-based delivery with synchronized feedback (Figure~\ref{fig:audio_feedback_mechanism}). The mechanism comprises three design principles:

\textbf{Pull-based audio delivery}: As introduced in §\ref{subsec:system_architecture}, the client requests each chunk only when ready for playback, and the Orchestrator serves from the Audio Buffer and tracks delivery accordingly. This synchronizes delivery with the client's playback rate, preventing buffer overflow.

\textbf{Synchronized token feedback}: Agent Speech Generator bundles each agent token with its corresponding audio in the messages it sends to the Orchestrator. The Orchestrator forwards agent tokens back to Agent Response Generator \emph{only after delivering to the client}. This ensures dialogue context contains only what users actually received.

\textbf{Interruption-safe design}: When barge-in occurs, the Orchestrator clears the undelivered Audio Buffer contents. Undelivered tokens never reach Agent Response Generator, maintaining context consistent with the user's actual experience. This design achieves a critical invariant that dialogue history precisely reflects the user's auditory experience, even during interruptions.

\section{Evaluation}
\label{sec:evaluation}

We evaluate semantic preservation, full-duplex interaction patterns, deployed-system latency, and explicit paralinguistic control. Model evaluations cover the two implementation paths from \S\ref{subsec:training_methodology},which we refer to as \textbf{TASTE2 (Merge)} and \textbf{TASTE2 (Direct)}, corresponding to Approach~1 and Approach~2, respectively. They share training data, objectives, and dialogue fine-tuning, differing only in how instruction-following capability is introduced.

\subsection{Semantic Preservation Analysis}
\label{subsec:semantic_preservation}

We first measure how much language-understanding accuracy remains after adapting a text language model into the TASTE2 Spoken LM and characterize that retention under both implementation paths.

\textbf{Benchmark Setup.} We adopt LLaMA-Questions (LLaMA-Q.)~\citep{benchmark-spoken-qa} as our Spoken QA benchmark, a question answering benchmark specifically designed for evaluating speech-text joint models. Following the evaluation protocol of the TASTE and Moshi, we employ answer containment accuracy as the evaluation metric to check whether the generated response contains the correct answer. To quantify how much semantic ability survives the adaptation, we compute the retention metric:
\begin{equation}
\text{Retention (\%)} = \frac{\text{SpokenLM}_{\text{Accuracy}}}{\text{LLM}_{\text{Accuracy}}} \times 100
\end{equation}
where 100\% means the spoken model matches its text reference, values above 100\% indicate an improvement over it, and higher values indicate better preservation.

\textbf{Experimental Configuration.} We evaluate both TASTE2 (Merge) and TASTE2 (Direct) on 300 questions from LLaMA-Questions after Stage SFT. Each spoken question is provided as audio, and the model jointly generates text tokens and audio latents. Accuracy relies on the same answer containment criterion mentioned above. This is applied after case normalization for both the ground truth answer and the generated response. The Spoken LM is evaluated in a zero shot setting, generating responses via greedy decoding. The text-only reference is Qwen2.5-7B Instruct, evaluated under the same zero-shot protocol. (Table~\ref{tab:semantic_retention}).

\begin{table}[H]
\centering
\caption{Semantic retention on the LLaMA-Questions benchmark. Retention is the spoken system's accuracy as a percentage of its own text reference; higher is better. T and S denote text and speech, respectively. Each spoken system is compared against the text-only model it was adapted from. TASTE2 uses Qwen2.5-7B Instruct, evaluated under the same zero-shot protocol; the other rows are taken from published setups with different backbones and prompting protocols, so their retention is not directly comparable.}
\label{tab:semantic_retention}
\begin{tabular}{lccc}
\toprule
\textbf{System} & \textbf{Mode} & \textbf{LLaMA-Q. Accuracy} & \textbf{Retention} \\
\midrule
Mini-Omni 0.5B(T$\xrightarrow{}$T) & T & 39.0\% & - \\
Mini-Omni 0.5B & T+S & 11.6\% & 29.7\% \\
\midrule
Helium 7B (text) & T & 75.0\% & - \\
Moshi 7B & T+S & 62.3\% & 83.1\% \\
\midrule
LLaMA3.1-8B-Instruct  & T & 71.7\% & - \\
Llama-Omni 8B & T+S & 67.3\% & 93.9\% \\
\midrule
Qwen2.5-7B Instruct & T & 57.3\% & - \\
\textbf{TASTE2 (Merge)} & T+S & 56.3\% & 98.2\% \\
\textbf{TASTE2 (Direct)} & T+S & 53.0\% & 92.4\% \\

\bottomrule
\end{tabular}
\end{table}

\textbf{Results and Analysis.} TASTE2 (Merge) preserves more semantic accuracy than TASTE2 (Direct). Two implications of this gap matter, both subject to the caveat that the other rows of Table~\ref{tab:semantic_retention} come from published setups with different backbones and prompting protocols.

The first concerns what the language model is asked to carry. The systems in Table~\ref{tab:semantic_retention} differ in how much acoustic responsibility sits inside the language model itself. Mini-Omni decodes text and audio in parallel streams, while Moshi predicts time-aligned text before audio across multiple codebook streams. The TASTE2 Spoken LM also emits acoustic information at every step, but its sequence is exactly as long as the text sequence, since one continuous latent travels alongside each text token instead of occupying positions of its own. Under that constraint it retains 98.2\% of the text reference. Llama-Omni also retains a large share, yet its language model emits text only and delegates acoustics to a downstream module, so its semantic accuracy is protected by construction rather than preserved under acoustic load. Text-aligned modeling thus enables per-token acoustic generation without the sequence length inflation that accompanies interleaved designs, and without the accuracy loss such designs report.

The second reading locates part of the remaining loss outside speech modeling. While TASTE2 (Merge) and TASTE2 (Direct) share the tokenizer, speech objectives, dialogue data, and Stage~SFT, they differ in how instruction-following capability is introduced. Ultimately, TASTE2 (Merge) retains 98.2\% against 92.4\% for TASTE2 (Direct). This indicates that a substantial share of the gap to the text reference stems from the instruction tuning method rather than speech representation, and can be avoided by using the model merging recipe.

\subsection{Residual VQ Ablation}
\label{subsec:ablation_rvq}

The Spoken LM predicts continuous audio latents rather than RVQ indices, making the need for quantization unclear. We therefore test whether RVQ acts as a useful training constraint.

\textbf{Experimental Setup.} We compare two training configurations: (1) \textbf{Without RVQ}: completely removing RVQ and directly training continuous latents; (2) \textbf{With RVQ (TASTE2)}: retaining RVQ as a discrete bottleneck. Both configurations use identical training data, hyperparameters, and model architecture except for RVQ presence. The training proceeds in two stages. Stage 1 separately trains each configuration's tokenizer and detokenizer for four epochs, after which we measure speech reconstruction accuracy (S3 accuracy). Stage 2 uses the respective tokenizers trained from Stage 1 to train the spoken language model for 50k steps, after which we measure latent space convergence via TASTE latent L1 distance and language modeling performance via text accuracy.

\begin{figure}[!ht]
\centering
\begin{tikzpicture}
\node at (0,0) [
    draw=black,
    fill=white,
    inner sep=0.3em,
    font=\small
] {
    \begin{tabular}{c@{\hspace{1em}}c}
    \tikz{\draw[blue,mark=square*,mark size=2pt,thick] plot coordinates {(0,0) (0.3,0)};} w/o RVQ (during Stage 2 training) &
    \tikz{\draw[red,mark=*,mark size=2pt,thick] plot coordinates {(0,0) (0.3,0)};} w/ RVQ (during Stage 2 training)
    \end{tabular}
};
\end{tikzpicture}

\vspace{0.2cm}

\begin{minipage}[t]{0.48\textwidth}
\centering
\begin{tikzpicture}
\begin{axis}[
    width=\textwidth,
    height=5.5cm,
    xlabel={Training Step},
    ylabel={Latent L1 Distance (log scale)},
    ylabel style={font=\small},
    ymode=log,
    xmin=10000, xmax=50000,
    xtick={10000,20000,30000,40000,50000},
    xticklabels={10k,20k,30k,40k,50k},
    scaled ticks=false,
    grid=major,
    tick label style={font=\scriptsize},
    label style={font=\small},
    title={TASTE Latent Space Convergence},
    title style={font=\small\bfseries}
]
\addplot[color=blue, mark=square, thick, mark size=2pt] coordinates {
    (10000, 29585.36)
    (20000, 21357.97)
    (30000, 17374.87)
    (40000, 15579.68)
    (50000, 14573.08)
};

\addplot[color=red, mark=*, thick, mark size=2pt] coordinates {
    (10000, 1.414)
    (20000, 1.392)
    (30000, 1.375)
    (40000, 1.362)
    (50000, 1.353)
};
\end{axis}
\end{tikzpicture}
\end{minipage}%
\hfill
\begin{minipage}[t]{0.48\textwidth}
\centering
\begin{tikzpicture}
\begin{axis}[
    width=\textwidth,
    height=5.5cm,
    xlabel={Training Step},
    ylabel={Text Accuracy},
    ylabel style={font=\small},
    xmin=10000, xmax=50000,
    ymin=0, ymax=0.45,
    xtick={10000,20000,30000,40000,50000},
    xticklabels={10k,20k,30k,40k,50k},
    scaled ticks=false,
    grid=major,
    tick label style={font=\scriptsize},
    label style={font=\small},
    title={Language Modeling Performance},
    title style={font=\small\bfseries}
]
\addplot[color=blue, mark=square, thick, mark size=2pt] coordinates {
    (10000, 0.0116)
    (20000, 0.0342)
    (30000, 0.0573)
    (40000, 0.0787)
    (50000, 0.0916)
};

\addplot[color=red, mark=*, thick, mark size=2pt] coordinates {
    (10000, 0.3814)
    (20000, 0.3840)
    (30000, 0.3828)
    (40000, 0.3852)
    (50000, 0.3867)
};
\end{axis}
\end{tikzpicture}
\end{minipage}

\caption{Effect of RVQ during two-stage training. Removing RVQ yields higher Stage~1 reconstruction accuracy but degrades Stage~2 latent-space convergence and language modeling performance. \textbf{Left:} Latent L1 distance over training. \textbf{Right:} Text accuracy over training.}
\label{fig:ablation_rvq}
\end{figure}
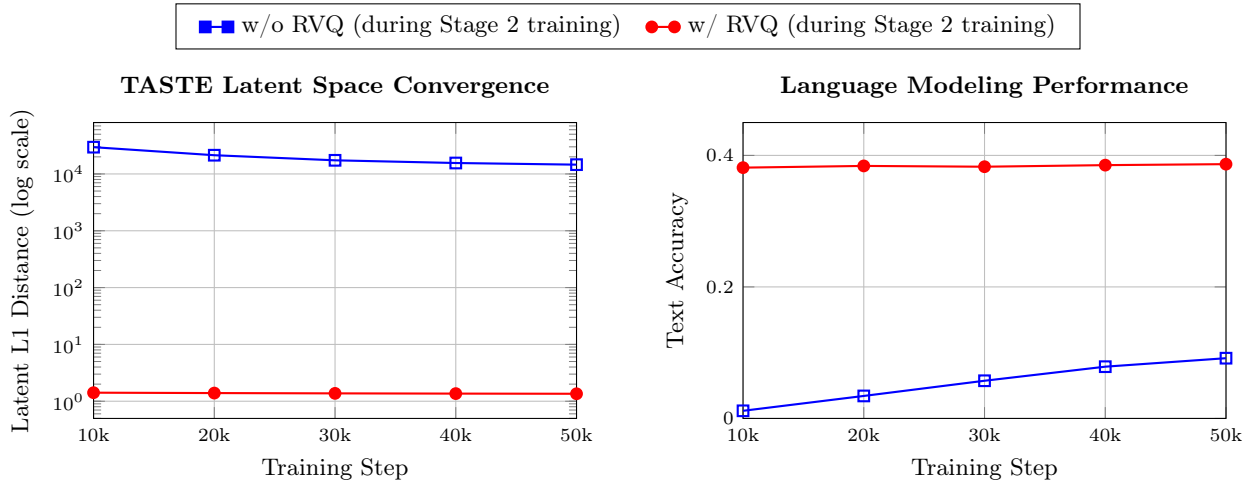

\textbf{Results and Analysis.} The two stages optimize for opposing constraints (Figure 5). Removing RVQ improves Stage 1 reconstruction (46.83\% versus 23.56\% S3 accuracy) because an unconstrained continuous space can flexibly encode the inputs required by the detokenizer. Yet, this same flexibility makes the space unlearnable in Stage 2. Latent L1 distance remains significantly higher (14,573 versus 1.35) and text accuracy falls to roughly 25\% of the RVQ model’s performance (9.16\% versus 38.67\%). This failure is not isolated to the acoustic branch. Since text tokens and audio latent are predicted in a single forward pass, an unachievable regression target ultimately degrades the text objective as well.

Consequently, RVQ does not inherently improve the fidelity of the representation. It simply renders the space predictable. This predictability is a structural requirement for the TASTE2 architecture and not just an incidental hyperparameter. A language model can successfully predict continuous acoustic latent only when quantization has adequately constrained the target space.

\subsection{Full-Duplex Interaction Evaluation}
\label{subsec:fullduplex_capability}

Full-Duplex-Bench~\citep{full-duplex-bench} evaluates whether the TASTE2 Spoken LM exhibits four interaction patterns, detailed below. These model-level tests are distinct from the deployed TASTE2 VoiceBot behavior described in §\ref{sec:system_design}.

\textbf{Benchmark Specification.} We use Full-Duplex-Bench v1.0, which assesses four interactive behaviors: (1) \emph{Pause Handling}: whether the model waits during user pauses rather than falsely starting to speak; (2) \emph{Backchannel}: whether the model provides brief listener feedback during user speech; (3) \emph{Smooth Turn Taking}: whether the model detects user turn-end and starts responding with low latency; and (4) \emph{User Interruption}: whether the model stops and adapts when a user barges in mid-response. The benchmark evaluates pause handling on both synthetic cases (Synth) and natural conversations drawn from the CANDOR corpus, and evaluates smooth turn taking only on the CANDOR cases. The benchmark employs automatic metrics for reproducible evaluation.

\textbf{Evaluation Metrics.} We adopt the metrics defined by Full-Duplex-Bench. \textbf{TOR (Takeover Rate)} measures how often the model takes over the speaking turn. For Pause Handling and Backchannel, lower TOR is better. It captures the model incorrectly claiming the turn during a user pause or in a context that calls for only a brief backchannel. This differs from \textbf{Frequency (Freq)} in Backchannel, which measures how often the model produces backchannels and a higher frequency means the model backchannels more like an engaged listener. \textbf{JSD} in Backchannel measures the Jensen--Shannon divergence between the model's predicted backchannel-timing distribution and the human-annotated ground-truth timing distribution. For Smooth Turn Taking and User Interruption, higher TOR is better because it reflects successful detection of turn-end or interruption. \textbf{Latency} is the time from the triggering event (user turn-end or interruption) to the start of the model's response, measured in seconds. \textbf{GPT-4o Score} evaluates the coherence and contextual appropriateness of responses after interruptions on a 1--5 scale.

\begin{table*}[!ht]
\centering
\caption{Full-Duplex-Bench v1.0 results across four interactive behaviors. Arrows indicate the preferred direction of each metric and bold marks the best value per column. Backchannel metrics are not reported for TASTE2 (\texttt{-}), whose deployed turn gate does not produce listener feedback. TOR stands for takeover rate. See \S\ref{subsec:fullduplex_capability} for metric details.}
\label{tab:fullduplex_results}
\small
\setlength{\tabcolsep}{3pt}
\resizebox{\textwidth}{!}{%
\begin{tabular}{lcccccccccc}
\toprule
\multirow{2}{*}{\textbf{Model}} & \multicolumn{2}{c}{\textbf{Pause Handling}} & \multicolumn{3}{c}{\textbf{Backchannel}} & \multicolumn{2}{c}{\textbf{Smooth Turn Taking}} & \multicolumn{3}{c}{\textbf{User Interruption}} \\
\cmidrule(lr){2-3} \cmidrule(lr){4-6} \cmidrule(lr){7-8} \cmidrule(lr){9-11}
& Synth TOR ↓ & CANDOR TOR ↓ & TOR ↓ & Freq ↑ & JSD ↓ & CANDOR TOR ↑ & Latency (s) ↓ & TOR ↑ & GPT-4o ↑ & Latency (s) ↓ \\
\midrule
dGSLM & 0.934 & 0.935 & 0.691 & 0.015 & 0.934 & \textbf{0.975} & 0.352 & 0.917 & 0.201 & 2.531 \\
Moshi & 0.985 & 0.980 & 1.000 & 0.001 & 0.957 & 0.941 & \textbf{0.265} & \textbf{1.000} & 0.765 & 0.257 \\
Freeze-Omni & 0.642 & 0.481 & 0.636 & 0.001 & 0.997 & 0.336 & 0.953 & 0.867 & \textbf{3.615} & 1.409 \\
Gemini Live & 0.255 & \textbf{0.310} & \textbf{0.091} & 0.012 & 0.896 & 0.655 & 1.301 & 0.891 & 3.376 & 1.183 \\
\midrule
TASTE2 (Merge) & \textbf{0.146} & 0.389 & - & - & - & 0.908 & 1.207 & \textbf{1.000} & 3.275 & \textbf{0.060} \\
TASTE2 (Direct) & 0.161 & 0.319 & - & - & - & 0.840 & 1.239 & 0.971 & 2.856 & 0.134 \\
\bottomrule
\end{tabular}%
}
\end{table*}

\textbf{Results and Analysis.} Table~\ref{tab:fullduplex_results} reveals a latency coherence trade-off. dGSLM and Moshi react quickly but fail to produce useful content after interruption, whereas Freeze-Omni and Gemini Live remain coherent but react slowly. TASTE2 (Merge) belongs to neither extreme. It stops on every interruption, reacts in 0.060~s, and still scores 3.275 on contextual appropriateness afterwards. Text-aligned modeling makes this combination possible because the dialogue state that survives an interruption is a text-token sequence that can be truncated exactly at the delivered speech boundary (§\ref{subsec:audio_playback_feedback}).

One weakness is unambiguous in both TASTE2 variants, which the models take over one second to make a smooth turn transition. This is the direct cost of the design. The turn decision waits for VAD speech end detection and then queries the token distribution used by model-gated turn decisions. This approach yields accurate turn identification without requiring a dedicated classifier, though it incurs a latency penalty. Comparing the two variants, TASTE2 (Merge) consistently outperforms TASTE2 (Direct) in handling user interruptions and smooth turn-taking. However, they exhibit a trade-off in pause handling: TASTE2 (Merge) achieves a better Synth TOR, while TASTE2 (Direct) performs better on the natural CANDOR conversations and remains near the best displayed system there.

\subsection{User Experience Latency Analysis}
\label{subsec:system_latency}

We measure the TASTE2 VoiceBot user experience latency using \textbf{Time To First Audio (TTFA)}, defined as the interval from the end of user speech to the onset of system speech. Unlike Full-Duplex-Bench v1.0, which evaluates pure backend latency, TTFA captures the complete end-to-end system delay by accounting for both backend processing and frontend connection overhead. We test at each of five input lengths: 4, 11, 25, 48, and 60 seconds. For each interaction, we locate the silence-to-voice transition in the generated waveform and verify it manually. TASTE2 VoiceBot runs with batch size~1 on two NVIDIA RTX A6000 GPUs. ChatGPT Voice, Gemini Live, and Grok were measured through their official mobile applications on March~13, 202 and their hardware configurations are unknown.

Table~\ref{tab:latency} presents TTFA measurements across systems. Because these five single-sample measurements show no consistent trend with input length, we summarize each system using its average TTFA. TensorRT, NVIDIA's inference optimization library, reduces TASTE2's average TTFA by 22\%, but TASTE2 still does not match ChatGPT Voice on this measure. We observe that the Agent Speech Generator as the primary end-to-end latency bottleneck. This suggest that making a lighter or more effective synthesis architecture is the most direct optimization target.

\begin{table}[!ht]
\centering
\caption{This table reports Time To First Audio (TTFA) in milliseconds across five input lengths. ChatGPT Voice, Gemini Live, and Grok were measured via their official mobile apps on March 13, 2026. TensorRT (TRT) lowers TASTE2's average TTFA from 3{,}469~ms to 2{,}701~ms, a 22\% reduction.}
\label{tab:latency}
\begin{tabular}{lcccccl}
\toprule
\textbf{System} & \multicolumn{6}{c}{\textbf{TTFA (ms)}} \\
\cmidrule(lr){2-7}
\multicolumn{1}{r}{\textit{Input Length}} & \textit{4s} & \textit{11s} & \textit{25s} & \textit{48s} & \textit{60s} & \textbf{Avg.} \\
\midrule
ChatGPT Voice  & 1056 & 1056 & 1120 & 1376 & 2240 & 1370 \\
Gemini Live & 4160 & 3968 & 3840 & 4800 & 3392 & 4032 \\
Grok & 2016 & 3424 & 1984 & 1952 & 2176 & 2310 \\
\hdashline
TASTE2 w/o TRT      & 3392 & 3232 & 3008 & 4000 & 3712 & 3469 \\
TASTE2            & 2464 & 2336 & 2912 & 3424 & 2368 & 2701 {\scriptsize(-22\%)} \\
\bottomrule
\end{tabular}
\end{table}

\subsection{Paralinguistic Control}
\label{subsec:paralinguistic_control}

\textbf{Setup.} We use two human rated diagnostics. First, the Stage~2 feature-continuation diagnostic establishes a baseline before doing any merging and SFT. We use the checkpoint before model merging and dialogue SFT and synthesize five unfinished English openings for each of eight features. The model will continue the aligned text--audio sequence, testing whether each feature can be continued at all before SFT strategy is applied. Second, we implement the explicit control diagnostic tests. After dialogue SFT, whether TASTE(Merge) and TASTE(Direct) can each produce the same eight features on instruction rather than merely continue them. We choose the 41 subset of VStyle's acoustic attributes~\citep{vstyle} evaluates instruction conditioned explicit control on both TASTE(Merge) and TASTE(Direct), with 10 \texttt{fast} and 10 \texttt{slow} Speed items, one \texttt{whisper} Volume item, and 4 items each for the five Emotion classes (\texttt{angry}, \texttt{fear}, \texttt{happy}, \texttt{sad}, \texttt{surprise}). Since the evaluation coverage differs from the full VStyle protocol, these scores cannot be directly compared with previously published VStyle results. In both diagnostics, three human raters independently evaluate every valid audio output. The raters assign a score ranging from 1 for completely inappropriate to 5 for perfectly appropriate. The final reported values are averaged across all raters and successful generations. Additionally, the Succ./Total columns in Table~\ref{tab:paralinguistic_features} report the number of attempts that successfully produced valid audio for each feature.

\begin{table}[!ht]
\centering
\caption{Human evaluations of paralinguistic feature continuation before dialogue SFT compared with style control after dialogue SFT. The rating protocol relies on three raters using a 1 to 5 scale and excludes invalid audio outputs (detailed in \S\ref{subsec:paralinguistic_control}). \textbf{Stage~2 feature continuation} applies five synthesized speech openings per feature. \textbf{VStyle explicit control} tests the subset of 41 items on both models after SFT. Values below 3 are highlighted in gray.}
\label{tab:paralinguistic_features}
\small
\renewcommand{\arraystretch}{1.08}
\begin{tabular}{lcccccc}
\toprule
& \multicolumn{2}{c}{\multirow{2}{*}{\textbf{Stage 2 Feature Continuation}}} & \multicolumn{4}{c}{\textbf{VStyle Explicit Control}} \\
\cmidrule(lr){4-7}
& \multicolumn{2}{c}{} & \multicolumn{2}{c}{\textbf{TASTE2 (Merge)}} & \multicolumn{2}{c}{\textbf{TASTE2 (Direct)}} \\
\cmidrule(lr){2-3} \cmidrule(lr){4-5} \cmidrule(lr){6-7}
\textbf{Feature} & Succ./Total & Avg $\pm$ Std & Succ./Total & Avg $\pm$ Std & Succ./Total & Avg $\pm$ Std \\
\midrule
fast     & 5/5 & $4.87 \pm 0.35$ & 9/10 & $4.15 \pm 0.95$ & 10/10 & $3.53 \pm 1.36$ \\
slow     & 5/5 & $4.27 \pm 0.96$ & 9/10 & $3.00 \pm 1.21$ & 10/10 & $3.77 \pm 1.25$ \\
happy    & 5/5 & $3.80 \pm 0.94$ & 4/4  & \textcolor[gray]{0.80}{$1.67 \pm 0.98$} & 3/4  & \textcolor[gray]{0.80}{$1.67 \pm 1.00$} \\
sad      & 3/5 & $3.56 \pm 1.24$ & 4/4  & \textcolor[gray]{0.80}{$2.00 \pm 1.35$} & 3/4  & \textcolor[gray]{0.80}{$1.89 \pm 1.05$} \\
surprise & 5/5 & $3.33 \pm 1.11$ & 4/4  & \textcolor[gray]{0.80}{$2.42 \pm 1.44$} & 4/4  & $3.42 \pm 1.56$ \\
whisper  & 4/5 & $3.27 \pm 1.19$ & 1/1  & \textcolor[gray]{0.80}{$1.00 \pm 0.00$} & 1/1  & \textcolor[gray]{0.80}{$1.00 \pm 0.00$} \\
fear     & 5/5 & \textcolor[gray]{0.80}{$2.40 \pm 1.24$} & 4/4  & \textcolor[gray]{0.80}{$1.17 \pm 0.58$} & 4/4  & \textcolor[gray]{0.80}{$1.33 \pm 0.65$} \\
angry    & 2/5 & \textcolor[gray]{0.80}{$2.00 \pm 1.26$} & 4/4  & \textcolor[gray]{0.80}{$1.33 \pm 0.65$} & 4/4  & $3.33 \pm 0.89$ \\
\bottomrule
\end{tabular}
\end{table}

\textbf{Results and Analysis.} Both TASTE2 models successfully produce valid audio for 39 of the 41 VStyle items after SFT. This shows the primary limitation lies in control rather than generation. To understand this bottleneck, we analyze performance progression from Stage 2 pretraining to the SFT stage.

The model demonstrates baseline representational capabilities in the Stage 2 continuation probe. The features \texttt{fast} and \texttt{slow} successfully clear the 3 point threshold, and \texttt{whisper} can also be effectively continued. However, attributes related to emotion such as \texttt{fear} and \texttt{angry} already exhibit weak performance at this early stage. After dialogue SFT, explicit control over these attributes diverges. While \texttt{fast} and \texttt{slow} remain robust under both TASTE(Direct) and TASTE(Merge) strategies, \texttt{whisper} can no longer be reliably produced on request despite its earlier viability. Furthermore, \texttt{fear} remains weak. The only notable exception is that emotion control exhibits strong strategy dependence. For instance, \texttt{angry} and \texttt{surprise} reach the threshold under the TASTE(Direct) strategy ($3.33 \pm 0.89$ and $3.42 \pm 1.56$ respectively) but fail under the TASTE(Merge) strategy ($1.33 \pm 0.65$ and $2.42 \pm 1.44$). The remaining attributes are generally weak or inconsistent.

Comparing these two stages reveals that Stage 2 pretraining acts as an upper bound for SFT performance in most cases. The quality of Stage 2 representations inherently constrains what a feature can achieve after SFT. If a paralinguistic attribute lacks a solid foundation in Stage 2, dialogue SFT alone is generally insufficient to establish robust control. Conversely, successful representation in Stage 2 does not guarantee survival through the SFT process. These patterns emphasize that high quality Stage 2 pretraining is critically important for enabling robust explicit control over speech attributes.

\subsection{Summary of Findings}
\label{subsec:summary_of_findings}

Taken together, the above evaluations answer the question raised in \S\ref{sec:introduction}, whether text-aligned speech tokenization is a viable modeling path toward full-duplex voice interaction, and they support a claim narrower and more specific than a general assertion of good performance.

\textbf{What the representation costs.} A language model that must also produce speech normally pays a penalty in sequence length. Because acoustic tokens occupy their own positions, the regularities learned during text pretraining become diluted across a longer and mixed interleaving sequence. The result shown in Table~\ref{tab:semantic_retention}. The interleaved systems retain between 29.7\% and 83.1\% of their text accuracy, in proportion to the space required by these acoustic positions. TASTE2 avoids this compromise entirely. By pairing one continuous audio latent with each text token, the model emits acoustic information at every step while keeping the overall sequence exactly as long as the text itself. As a result, it retains 98.2\% of its text reference.

\textbf{Where it pays off.} The primary advantage emerges during full duplex turn boundaries such as interruptions, where TASTE2 reacts in just 0.060~s while maintaining ability in coherence and contextual
appropriateness. Other evaluated systems fail to balance this rapid speed with coherent generation. This capability stems directly from the underlying text representation, which allows the dialogue history to be truncated exactly at the interruption point so the model resumes precisely from what the user actually heard. Turn identification benefits similarly because a turn boundary acts as an ordinary token, enabling TOR to reach 0.908 without requiring any dedicated classifier.

\textbf{What it does not yet deliver.} Routing turn decisions through the model distribution improves accuracy but increases latency, resulting in a smooth turn latency of 1.207~s compared to 0.265~s for the fastest evaluated system. The bottleneck is in the Agent Speech Generator rather than the Spoken LM. Furthermore, explicit paralinguistic control remains inconsistent across attributes and training strategies. The \texttt{fast} and \texttt{slow} features are the only ones to clear after SFT the models, while other emotion features remain weak.

\section{Conclusion}
\label{sec:conclusion}

TASTE2 establishes TASTE based text aligned speech modeling and deployment as a viable path toward full duplex voice interaction. Achieving this required an end to end adaptation of the original utterance level method. A shared text token vocabulary eliminates word level averaging, and pairing one continuous audio latent with each text token allows the model to emit acoustic information at every step without extending the text sequence length. Interruption aware dialogue training teaches the model when to yield or take a turn, while the Speech Detokenizer produces S3 units incrementally. The TASTE2 VoiceBot integrates this model into a working research deployment with VAD-triggered and model-gated turn decisions, streaming synthesis, and barge-in handling.

The evaluations (\S\ref{subsec:summary_of_findings}) support a specific claim rather than a general one. Text-aligned speech tokenization preserves text-model competence at a measurable cost and delivers the full-duplex behavior that depends most directly on that representation. Turn-taking latency, deployed synthesis speed, and cross-attribute paralinguistic control remain open limits.

\section*{Acknowledgments}

The authors gratefully acknowledge the NVIDIA AI Technology Center (NVAITC) for providing access to the Taipei-1 supercomputer and the computational resources used in this research.

\bibliographystyle{plainnat}
\bibliography{references}

@inproceedings{taste,
  title={Taste: Text-aligned speech tokenization and embedding for spoken language modeling},
  author={Tseng, Liang-Hsuan and Chen, Yi-Chang and Lee, Kuan and Shiu, Da-Shan and Lee, Hung-yi},
  booktitle={International Conference on Learning Representations},
  volume={2026},
  pages={125293--125314},
  year={2026}
}

@article{gslm,
  title={On generative spoken language modeling from raw audio},
  author={Lakhotia, Kushal and Kharitonov, Eugene and Hsu, Wei-Ning and Adi, Yossi and Polyak, Adam and Bolte, Benjamin and Nguyen, Tu-Anh and Copet, Jade and Baevski, Alexei and Mohamed, Abdelrahman and others},
  journal={Transactions of the Association for Computational Linguistics},
  volume={9},
  pages={1336--1354},
  year={2021},
  publisher={MIT Press One Rogers Street, Cambridge, MA 02142-1209, USA journals-info~…}
}

@article{dgslm,
  title={Generative spoken dialogue language modeling},
  author={Nguyen, Tu Anh and Kharitonov, Eugene and Copet, Jade and Adi, Yossi and Hsu, Wei-Ning and Elkahky, Ali and Tomasello, Paden and Algayres, Robin and Sagot, Benoit and Mohamed, Abdelrahman and others},
  journal={Transactions of the Association for Computational Linguistics},
  volume={11},
  pages={250--266},
  year={2023},
  publisher={MIT Press One Broadway, 12th Floor, Cambridge, Massachusetts 02142, USA~…}
}

@article{spiritlm,
  title={Spirit-lm: Interleaved spoken and written language model},
  author={Nguyen, Tu Anh and Muller, Benjamin and Yu, Bokai and Costa-Jussa, Marta R and Elbayad, Maha and Popuri, Sravya and Ropers, Christophe and Duquenne, Paul-Ambroise and Algayres, Robin and Mavlyutov, Ruslan and others},
  journal={Transactions of the Association for Computational Linguistics},
  volume={13},
  pages={30--52},
  year={2025},
  publisher={MIT Press 255 Main Street, 9th Floor, Cambridge, Massachusetts 02142, USA~…}
}

@article{moshi,
  title={Moshi: a speech-text foundation model for real-time dialogue},
  author={D{\'e}fossez, Alexandre and Mazar{\'e}, Laurent and Orsini, Manu and Royer, Am{\'e}lie and P{\'e}rez, Patrick and J{\'e}gou, Herv{\'e} and Grave, Edouard and Zeghidour, Neil},
  journal={arXiv preprint arXiv:2410.00037},
  year={2024}
}

@article{mini-omni,
  title={Mini-omni: Language models can hear, talk while thinking in streaming},
  author={Xie, Zhifei and Wu, Changqiao},
  journal={arXiv preprint arXiv:2408.16725},
  year={2024}
}

@inproceedings{llama-omni,
  title={Llama-omni: Seamless speech interaction with large language models},
  author={Fang, Qingkai and Guo, Shoutao and Zhou, Yan and Ma, Zhengrui and Zhang, Shaolei and Feng, Yang},
  booktitle={International Conference on Learning Representations},
  volume={2025},
  pages={57607--57624},
  year={2025}
}

@inproceedings{llama-omni2,
  title={LLaMA-omni 2: LLM-based real-time spoken chatbot with autoregressive streaming speech synthesis},
  author={Fang, Qingkai and Zhou, Yan and Guo, Shoutao and Zhang, Shaolei and Feng, Yang},
  booktitle={Proceedings of the 63rd Annual Meeting of the Association for Computational Linguistics (Volume 1: Long Papers)},
  pages={18617--18629},
  year={2025}
}

@article{glm4-voice,
  title={Glm-4-voice: Towards intelligent and human-like end-to-end spoken chatbot},
  author={Zeng, Aohan and Du, Zhengxiao and Liu, Mingdao and Wang, Kedong and Jiang, Shengmin and Zhao, Lei and Dong, Yuxiao and Tang, Jie},
  journal={arXiv preprint arXiv:2412.02612},
  year={2024}
}

@article{kimi-audio,
  title={Kimi-audio technical report},
  author={Ding, Ding and Ju, Zeqian and Leng, Yichong and Liu, Songxiang and Liu, Tong and Shang, Zeyu and Shen, Kai and Song, Wei and Tan, Xu and Tang, Heyi and others},
  journal={arXiv preprint arXiv:2504.18425},
  year={2025}
}

@article{baichuan-audio,
  title={Baichuan-audio: A unified framework for end-to-end speech interaction},
  author={Li, Tianpeng and Liu, Jun and Zhang, Tao and Fang, Yuanbo and Pan, Da and Wang, Mingrui and Liang, Zheng and Li, Zehuan and Lin, Mingan and Dong, Guosheng and others},
  journal={arXiv preprint arXiv:2502.17239},
  year={2025}
}

@misc{qwen2-omni,
      title={Qwen2.5-Omni Technical Report}, 
      author={Jin Xu and Zhifang Guo and Jinzheng He and Hangrui Hu and Ting He and Shuai Bai and Keqin Chen and Jialin Wang and Yang Fan and Kai Dang and Bin Zhang and Xiong Wang and Yunfei Chu and Junyang Lin},
      year={2025},
      eprint={2503.20215},
      archivePrefix={arXiv},
      primaryClass={cs.CL},
      url={https://arxiv.org/abs/2503.20215}, 
}

@article{freeze-omni,
  title={Freeze-omni: A smart and low latency speech-to-speech dialogue model with frozen llm},
  author={Wang, Xiong and Li, Yangze and Fu, Chaoyou and Shen, Yunhang and Xie, Lei and Li, Ke and Sun, Xing and Ma, Long},
  journal={arXiv preprint arXiv:2411.00774},
  year={2024}
}

@article{think-before-talk,
  title={Think before you talk: Enhancing meaningful dialogue generation in full-duplex speech language models with planning-inspired text guidance},
  author={Cui, Wenqian and Zhu, Lei and Li, Xiaohui and Guo, Zhihan and Bai, Haoli and Hou, Lu and King, Irwin},
  journal={arXiv e-prints},
  pages={arXiv--2508},
  year={2025}
}

@article{salmonn-omni,
  title={Salmonn-omni: A codec-free llm for full-duplex speech understanding and generation},
  author={Yu, Wenyi and Wang, Siyin and Yang, Xiaoyu and Chen, Xianzhao and Tian, Xiaohai and Zhang, Jun and Sun, Guangzhi and Lu, Lu and Wang, Yuxuan and Zhang, Chao},
  journal={arXiv preprint arXiv:2411.18138},
  year={2024}
}

@article{salm-duplex,
  title={Salm-duplex: Efficient and direct duplex modeling for speech-to-speech language model},
  author={Hu, Ke and Hosseini-Asl, Ehsan and Chen, Chen and Casanova, Edresson and Ghosh, Subhankar and {\.Z}elasko, Piotr and Chen, Zhehuai and Li, Jason and Balam, Jagadeesh and Ginsburg, Boris},
  journal={arXiv preprint arXiv:2505.15670},
  year={2025}
}

@article{ntpp,
  title={Ntpp: Generative speech language modeling for dual-channel spoken dialogue via next-token-pair prediction},
  author={Wang, Qichao and Meng, Ziqiao and Cui, Wenqian and Zhang, Yifei and Wu, Pengcheng and Wu, Bingzhe and King, Irwin and Chen, Liang and Zhao, Peilin},
  journal={arXiv preprint arXiv:2506.00975},
  year={2025}
}

@article{prosodylm,
  title={ProsodyLM: Uncovering the emerging prosody processing capabilities in speech language models},
  author={Qian, Kaizhi and Fan, Xulin and Ni, Junrui and Shechtman, Slava and Hasegawa-Johnson, Mark and Gan, Chuang and Zhang, Yang},
  journal={arXiv preprint arXiv:2507.20091},
  year={2025}
}

@inproceedings{paras2s,
  title={Paras2s: Benchmarking and aligning spoken language models for paralinguistic-aware speech-to-speech interaction},
  author={Yang, Shu-wen and Tu, Ming and Liu, Ting-Wei and Qu, Xinghua and Lee, Hung-yi and Lu, Lu and Wang, Yuxuan and Wu, Yonghui},
  booktitle={International Conference on Learning Representations},
  volume={2026},
  pages={70704--70729},
  year={2026}
}

@inproceedings{emo-reasoning,
  title={Emo-reasoning: Benchmarking emotional reasoning capabilities in spoken dialogue systems},
  author={Liu, Jingwen and Cheng, Kan Jen and Lian, Jiachen and Anand, Akshay and Jain, Rishi and Qiao, Faith and Netzorg, Robin and Chou, Huang-Cheng and Li, Tingle and Anumanchipalli, Gopala and others},
  booktitle={2025 IEEE Automatic Speech Recognition and Understanding Workshop (ASRU)},
  pages={1--8},
  year={2025},
  organization={IEEE}
}

@article{cosyvoice,
  title={Cosyvoice: A scalable multilingual zero-shot text-to-speech synthesizer based on supervised semantic tokens},
  author={Du, Zhihao and Chen, Qian and Zhang, Shiliang and Hu, Kai and Lu, Heng and Yang, Yexin and Hu, Hangrui and Zheng, Siqi and Gu, Yue and Ma, Ziyang and others},
  journal={arXiv preprint arXiv:2407.05407},
  year={2024}
}

@article{cosyvoice2,
  title={Cosyvoice 2: Scalable streaming speech synthesis with large language models},
  author={Du, Zhihao and Wang, Yuxuan and Chen, Qian and Shi, Xian and Lv, Xiang and Zhao, Tianyu and Gao, Zhifu and Yang, Yexin and Gao, Changfeng and Wang, Hui and others},
  journal={arXiv preprint arXiv:2412.10117},
  year={2024}
}

@article{voicebench,
  title={Voicebench: Benchmarking llm-based voice assistants},
  author={Chen, Yiming and Yue, Xianghu and Zhang, Chen and Gao, Xiaoxue and Tan, Robby T and Li, Haizhou},
  journal={URL https://arxiv. org/abs/2410},
  volume={17196},
  year={2024}
}

@inproceedings{audiobench,
  title={Audiobench: A universal benchmark for audio large language models},
  author={Wang, Bin and Zou, Xunlong and Lin, Geyu and Sun, Shuo and Liu, Zhuohan and Zhang, Wenyu and Liu, Zhengyuan and Aw, AiTi and Chen, Nancy},
  booktitle={Proceedings of the 2025 Conference of the Nations of the Americas Chapter of the Association for Computational Linguistics: Human Language Technologies (Volume 1: Long Papers)},
  pages={4297--4316},
  year={2025}
}

@inproceedings{full-duplex-bench-v15,
  title={Full-duplex-bench v1. 5: Evaluating overlap handling for full-duplex speech models},
  author={Lin, Guan-Ting and Kuan, Shih-Yun Shan and Wang, Qirui and Lian, Jiachen and Li, Tingle and Watanabe, Shinji and Lee, Hung-yi},
  booktitle={ICASSP 2026-2026 IEEE International Conference on Acoustics, Speech and Signal Processing (ICASSP)},
  pages={19447--19451},
  year={2026},
  organization={IEEE}
}

@article{fd-bench,
  title={Fd-bench: A full-duplex benchmarking pipeline designed for full duplex spoken dialogue systems},
  author={Peng, Yizhou and Chao, Yi-Wen and Ng, Dianwen and Ma, Yukun and Ni, Chongjia and Ma, Bin and Chng, Eng Siong},
  journal={arXiv preprint arXiv:2507.19040},
  year={2025}
}

@inproceedings{full-duplex-bench,
  title={Full-duplex-bench: A benchmark to evaluate full-duplex spoken dialogue models on turn-taking capabilities},
  author={Lin, Guan-Ting and Lian, Jiachen and Li, Tingle and Wang, Qirui and Anumanchipalli, Gopala and Liu, Alexander H and Lee, Hung-yi},
  booktitle={2025 IEEE Automatic Speech Recognition and Understanding Workshop (ASRU)},
  pages={1--8},
  year={2025},
  organization={IEEE}
}

@inproceedings{talking-turns,
  title={Talking turns: Benchmarking audio foundation models on turn-taking dynamics},
  author={Arora, Siddhant and Lu, Zhiyun and Chiu, Chung-Cheng and Pang, Ruoming and Watanabe, Shinji},
  booktitle={International Conference on Learning Representations},
  volume={2025},
  pages={52754--52781},
  year={2025}
}

@article{televal,
  title={Televal: A dynamic benchmark designed for spoken language models in chinese interactive scenarios},
  author={Li, Zehan and Chen, Hongjie and Wang, Qing and Zhang, Yuxin and Zhou, Jing and Lv, Hang and Du, Mengjie and Song, Yaodong and Lian, Jie and Kang, Jian and others},
  journal={arXiv preprint arXiv:2507.18061},
  year={2025}
}

@inproceedings{whisper,
  title={Robust speech recognition via large-scale weak supervision},
  author={Radford, Alec and Kim, Jong Wook and Xu, Tao and Brockman, Greg and McLeavey, Christine and Sutskever, Ilya},
  booktitle={International conference on machine learning},
  pages={28492--28518},
  year={2023},
  organization={PMLR}
}

@article{hubert,
  title={Hubert: Self-supervised speech representation learning by masked prediction of hidden units},
  author={Hsu, Wei-Ning and Bolte, Benjamin and Tsai, Yao-Hung Hubert and Lakhotia, Kushal and Salakhutdinov, Ruslan and Mohamed, Abdelrahman},
  journal={IEEE/ACM transactions on audio, speech, and language processing},
  volume={29},
  pages={3451--3460},
  year={2021},
  publisher={IEEE}
}

@article{soundstream,
  author={Zeghidour, Neil and Luebs, Alejandro and Omran, Ahmed and Skoglund, Jan and Tagliasacchi, Marco},
  journal={IEEE/ACM Transactions on Audio, Speech, and Language Processing}, 
  title={SoundStream: An End-to-End Neural Audio Codec}, 
  year={2022},
  volume={30},
  number={},
  pages={495-507},
  doi={10.1109/TASLP.2021.3129994}}

@inproceedings{speechtokenizer,
  title={Speechtokenizer: Unified speech tokenizer for speech language models},
  author={Zhang, Xin and Zhang, Dong and Li, Shimin and Zhou, Yaqian and Qiu, Xipeng},
  booktitle={International Conference on Learning Representations},
  volume={2024},
  pages={31798--31818},
  year={2024}
}

@article{audiolm,
  title={Audiolm: a language modeling approach to audio generation},
  author={Borsos, Zal{\'a}n and Marinier, Rapha{\"e}l and Vincent, Damien and Kharitonov, Eugene and Pietquin, Olivier and Sharifi, Matt and Roblek, Dominik and Teboul, Olivier and Grangier, David and Tagliasacchi, Marco and others},
  journal={IEEE/ACM transactions on audio, speech, and language processing},
  volume={31},
  pages={2523--2533},
  year={2023},
  publisher={IEEE}
}

@inproceedings{speechgpt,
  title={Speechgpt: Empowering large language models with intrinsic cross-modal conversational abilities},
  author={Zhang, Dong and Li, Shimin and Zhang, Xin and Zhan, Jun and Wang, Pengyu and Zhou, Yaqian and Qiu, Xipeng},
  booktitle={Findings of the Association for Computational Linguistics: EMNLP 2023},
  pages={15757--15773},
  year={2023}
}

@article{audiopalm,
  title={Audiopalm: A large language model that can speak and listen (2023)},
  author={Rubenstein, Paul K and Asawaroengchai, Chulayuth and Nguyen, Duc Dung and Bapna, Ankur and Borsos, Zal{\'a}n and de Chaumont Quitry, F{\'e}lix and Chen, Peter and El Badawy, Dalia and Han, Wei and Kharitonov, Eugene and others},
  journal={arXiv preprint arXiv:2306.12925},
  year={2023}
}

@article{qwen-audio,
  title={Qwen-audio: Advancing universal audio understanding via unified large-scale audio-language models},
  author={Chu, Yunfei and Xu, Jin and Zhou, Xiaohuan and Yang, Qian and Zhang, Shiliang and Yan, Zhijie and Zhou, Chang and Zhou, Jingren},
  journal={arXiv preprint arXiv:2311.07919},
  year={2023}
}

@inproceedings{salmonn,
  title={Salmonn: Towards generic hearing abilities for large language models},
  author={Tang, Changli and Yu, Wenyi and Sun, Guangzhi and Chen, Xianzhao and Tan, Tian and Li, Wei and Lu, Lu and Ma, Zejun and Zhang, Chao},
  booktitle={International Conference on Learning Representations},
  volume={2024},
  pages={16607--16629},
  year={2024}
}

@inproceedings{syncllm,
  title={Beyond turn-based interfaces: Synchronous llms as full-duplex dialogue agents},
  author={Veluri, Bandhav and Peloquin, Benjamin N and Yu, Bokai and Gong, Hongyu and Gollakota, Shyamnath},
  booktitle={Proceedings of the 2024 Conference on Empirical Methods in Natural Language Processing},
  pages={21390--21402},
  year={2024}
}

@inproceedings{omniflatten,
  title={Omniflatten: An end-to-end gpt model for seamless voice conversation},
  author={Zhang, Qinglin and Cheng, Luyao and Deng, Chong and Chen, Qian and Wang, Wen and Zheng, Siqi and Liu, Jiaqing and Yu, Hai and Tan, Chao-Hong and Du, Zhihao and others},
  booktitle={Proceedings of the 63rd Annual Meeting of the Association for Computational Linguistics (Volume 1: Long Papers)},
  pages={14570--14580},
  year={2025}
}

@article{fd-slm-survey,
  title={From turn-taking to synchronous dialogue: A survey of full-duplex spoken language models},
  author={Chen, Yuxuan and Yu, Haoyuan},
  journal={arXiv preprint arXiv:2509.14515},
  year={2025}
}

@article{wavchat,
  title={Wavchat: A survey of spoken dialogue models},
  author={Ji, Shengpeng and Chen, Yifu and Fang, Minghui and Zuo, Jialong and Lu, Jingyu and Wang, Hanting and Jiang, Ziyue and Zhou, Long and Liu, Shujie and Cheng, Xize and others},
  journal={arXiv preprint arXiv:2411.13577},
  year={2024}
}

@inproceedings{turngpt,
  title={TurnGPT: a transformer-based language model for predicting turn-taking in spoken dialog},
  author={Ekstedt, Erik and Skantze, Gabriel},
  booktitle={Findings of the Association for Computational Linguistics: EMNLP 2020},
  pages={2981--2990},
  year={2020}
}

@inproceedings{vap,
  title={{Voice Activity Projection: Self-supervised Learning of Turn-taking Events}},
  author={Ekstedt, Erik and Skantze, Gabriel},
  booktitle={Proceedings of Interspeech},
  year={2022},
  url={https://arxiv.org/abs/2205.09812}
}

@inproceedings{benchmark-spoken-qa,
  title={Spoken question answering and speech continuation using spectrogram-powered llm},
  author={Nachmani, Eliya and Levkovitch, Alon and Hirsch, Roy and Salazar, Julian and Asawaroengchai, Chulayuth and Mariooryad, Soroosh and Rivlin, Ehud and Skerry-Ryan, RJ and Tadmor Ramanovich, Michele},
  booktitle={International Conference on Learning Representations},
  volume={2024},
  pages={51883--51898},
  year={2024}
}

@inproceedings{emilia,
  title={Emilia: An extensive, multilingual, and diverse speech dataset for large-scale speech generation},
  author={He, Haorui and Shang, Zengqiang and Wang, Chaoren and Li, Xuyuan and Gu, Yicheng and Hua, Hua and Liu, Liwei and Yang, Chen and Li, Jiaqi and Shi, Peiyang and others},
  booktitle={2024 IEEE Spoken Language Technology Workshop (SLT)},
  pages={885--890},
  year={2024},
  organization={IEEE}
}

@inproceedings{libritts,
  title={{LibriTTS: A Corpus Derived from LibriSpeech for Text-to-Speech}},
  author={Zen, Heiga and Dang, Viet and Clark, Rob and Zhang, Yu and Weiss, Ron J and Jia, Ye and Chen, Zhifeng and Wu, Yonghui},
  booktitle={Proceedings of Interspeech},
  pages={1526--1530},
  year={2019},
  url={https://arxiv.org/abs/1904.02882}
}

@inproceedings{vstyle,
  title={Vstyle: A benchmark for voice style adaptation with spoken instructions},
  author={Zhan, Jun and Han, Mingyang and Xie, Yuxuan and Wang, Chen and Zhang, Dong and Huang, Kexin and Shi, Haoxiang and Wang, DongXiao and Song, Tengtao and Cheng, Qinyuan and others},
  booktitle={ICASSP 2026-2026 IEEE International Conference on Acoustics, Speech and Signal Processing (ICASSP)},
  pages={19477--19481},
  year={2026},
  organization={IEEE}
}

@inproceedings{ties,
  title={TIES-Merging: Resolving Interference When Merging Models},
  author={Yadav, Prateek and Tam, Derek and Choshen, Leshem and Raffel, Colin and Bansal, Mohit},
  booktitle={Advances in Neural Information Processing Systems},
  volume={36},
  pages={15886--15911},
  year={2023},
  url={https://arxiv.org/abs/2306.01708}
}

@article{speechparaling-bench,
  title={SpeechParaling-Bench: A Comprehensive Benchmark for Paralinguistic-Aware Speech Generation},
  author={Liu, Ruohan and Yin, Shukang and Wang, Tao and Zhang, Dong and Zhuang, Weiji and Ren, Shuhuai and He, Ran and Shan, Caifeng and Fu, Chaoyou},
  journal={arXiv preprint arXiv:2604.20842},
  year={2026}
}

@inproceedings{echomind,
  title={Echomind: An interrelated multi-level benchmark for evaluating empathetic speech language models},
  author={Zhou, Li and Yu, Lutong and Lyu, You and Lin, Yihang and Zhao, Zefeng and Ao, Junyi and Zhang, Yuhao and Benyou, Wang and Li, Haizhou},
  booktitle={International Conference on Learning Representations},
  volume={2026},
  pages={55554--55588},
  year={2026}
}

@article{encodec,
  title={High fidelity neural audio compression},
  author={D{\'e}fossez, Alexandre and Copet, Jade and Synnaeve, Gabriel and Adi, Yossi},
  journal={arXiv preprint arXiv:2210.13438},
  year={2022}
}

\end{document}